\documentclass[useAMS,usenatbib]{mn2e}
\usepackage{graphicx,amssymb}
\newcommand{\bc}{\begin{center}}
\newcommand{\ec}{\end{center}}

\title[The Galaxy and its stellar halo - a hybrid cosmological approach]
      {The Galaxy and its stellar halo: insights on their formation from a
      hybrid cosmological approach}   
\author[G.~De Lucia \& A. Helmi]
       {Gabriella De Lucia$^1$\thanks{Email: gdelucia@mpa-garching.mpg.de}
         and Amina Helmi$^2$
         \\      
        $^1$Max--Planck--Institut f\"ur Astrophysik, 
        Karl--Schwarzschild--Str. 1, D-85748 Garching, Germany\\
        $^2$Kapteyn Astronomical Institute, University of Groningen, 
        P.O. Box 800, 9700 AV Groningen, Netherlands}
\begin{document}

\pagerange{\pageref{firstpage}--\pageref{lastpage}} 
\pubyear{2008}

\maketitle

\label{firstpage}

\begin{abstract}
  We use a series of high-resolution simulations of a `Milky-Way' halo coupled
  to semi-analytic methods to study the formation of our own Galaxy and of its
  stellar halo. The physical properties of our model Milky Way, as well as the
  age and metallicity distribution of stars in the different components, are in
  relatively good agreement with observational measurements. Assuming that the
  stellar halo builds up from the cores of the satellite galaxies that merged
  with the Milky Way over its life-time, we are able to study the physical and
  structural properties of this component. In agreement with previous work, we
  find that the largest contribution to the stellar halo should come from a few
  relatively massive ($10^8-10^{10}\,{\rm M}_{\sun}$) satellites, accreted at
  early times. Our ``stellar halo'' does not exhibit any clear metallicity
  gradient, but higher metallicity stars are more centrally concentrated than
  stars of lower abundance. This indicates that the probability of observing
  low-metallicity halo stars increases with distance from the Galactic centre.
  We find that the proposed ``dual'' nature of the Galactic stellar halo can be
  explained in our model as a result of a mass-metallicity relation imprinted
  in the building blocks of this component.
\end{abstract}


\section{Introduction}
\label{sec:intro}

Our own galaxy - the Milky Way - is a fairly large spiral galaxy consisting of
four main stellar components. Most of the stars are distributed in a thin disk,
exhibit a wide range of ages, and are on high angular momentum orbits. A much
smaller mass of stars (about 10-20 per cent of that in the thin disk) reside in
a distinct component which was established about 25 years ago though star
counts \citep{Gilmore_and_Reid_1983}, and which is referred to as the `thick
disk'. The stars in the thick disk are old, have on average lower metallicity
than those of similar age in the thin disk, and are on orbits of lower angular
momentum. The Galactic bulge is dominated by an old and relatively metal-rich
stellar population with a tail to low abundances. It has a peanut-shape
\citep{Dwek_etal_1995}, is kinematically hotter than the Milky Way disk but
colder than the Milky Way halo \citep{Kuijken_and_Rich_2002}. The stellar halo
represents only a tiny fraction of the total stellar mass ($\sim
2\times10^9\,{\rm M}_{\odot}$ - \citealt*{Carney_Latham_and_Laird_1989}), and
is dominated by old and metal poor stars which reside on low angular momentum
orbits.

While the Milky Way is only one galaxy, it is the one that we can study in
unique detail. Accurate measurements of ages, metallicities, and kinematics
have been collected over the years for a large number of individual stars. Over
the next decade, a number of astrometric and spectroscopic surveys will provide
accurate spatial, kinematic, and chemical information for a much larger number
of stars (e.g. the satellite {\it Gaia} - \citealt{Perryman_etal_2001};
the Radial Velocity Experiment {\it RAVE} - \citealt{Steinmetz_etal_2006};
and the Sloan Extension for Galactic Understanding and Exploration {\it SEGUE}
- \citealt{Beers_etal_2004}). This vast amount of fossil information will
provide important advances in our understanding of the sequence of events which
led to the formation of our Galaxy \citep{Freeman_Bland-Hawthorn_2002}.

Historically, chemical and kinematic information were used as a basis to
formulate the first galaxy formation models. In their classical paper,
\citet*{Eggen_Lynden-Bell_Sandage_1962} analysed the properties and motion of
221 dwarfs and showed that those of lower metallicity tended to move on more
highly eccentric orbits. The observed trends were interpreted as a signature
that the stars now observed as a spheroidal halo formed during a rapid radial
collapse that later continued to form the stellar disk. About one decade later,
\citet{Searle_and_Zinn_1978} measured the metallicities in a sample of globular
clusters and found no significant abundance gradient as a function of the
galactocentric distance. These observations led
\citeauthor{Searle_and_Zinn_1978} to formulate the hypothesis that the stellar
halo formed over a longer timescale through agglomeration of many sub-galactic
`fragments' that may be similar to the surviving dwarf spheroidal galaxies
today observed as satellites of the Milky Way. The observational evidence in
support of this scenario has mounted significantly in the last decade ranging
from detection of significant clumpiness in the phase space distribution of
halo and disk stars (e.g.  \citealt*{Majewski_Munn_and_Hawley_1996};
\citealt{Helmi_etal_1999}; \citealt{Chiba_and_Beers_2000};
\citealt{Helmi_etal_2006a}) to the direct detection of satellite galaxies
caught in the act of tidal disruption (e.g.  \citealt*{Ibata_etal_1994};
\citealt{Martin_etal_2004}; \citealt{Zucker_etal_2006}).

The \citeauthor{Searle_and_Zinn_1978} scenario, and the observational results
mentioned above, are in qualitative agreement with expectations from the
hierarchical cold dark matter (CDM) model which, with the `concordance' set of
cosmological parameters ($\Lambda$CDM), boasts a considerable degree of success
in reproducing a large number of observations at various cosmic epochs, in
particular on large scales. On galactic and sub-galactic scales, the success of
the model has not been convincingly demonstrated yet, and a number of issues
remain subject of a lively debate in the astronomical community. Among these
are the dwarf galaxy counts, the degree of concentration of dark matter haloes,
and the rotation curves of dwarfs and low surface brightness galaxies
\citep[e.g.][and references
therein]{Simon_etal_2003,Kazantzidis_etal_2004,Stoehr2006,Strigari_etal_2007}.

The origin and structure of the stellar halo has been studied by several
authors, using a variety of techniques \citep[for a review,
  see][]{Helmi_2008}. These include cosmological numerical simulations with and
without baryonic physics
\citep*[e.g.][]{Diemand_Madau_Moore_2005,Abadi_etal_2006}, and phenomenological
modelling of the evolution of baryons inside haloes, usually in combination
with N-body simulations that provide the dynamical history of the system
\citep{Bullock_Johnston_2005}.

A number of recent studies have raised concerns about the
\citeauthor{Searle_and_Zinn_1978} scenario on the basis of the observation that
stars in Local Group dwarf spheroidals (dSph's) tend to have lower $\alpha$
abundances than stars in the stellar halo
\citep{Shetrone_etal_2001,Tolstoy_etal_2003,Venn_etal_2004}. The observed
abundance pattern excludes the possibility that a significant contribution to
the stellar halo comes from disrupted satellites similar to the {\it present
  day} dSph's. It is not entirely unexpected that the surviving satellites
might be intrinsically different from the main contributors to the stellar
halo, given they had a Hubble time to evolve as independent entities.  This
argument has been put forward in a recent series of papers by
\citet{Bullock_Johnston_2005}, \citet{Robertson_etal_2005}, and
\citet{Font_etal_2006} who reproduce the observed chemical abundance pattern by
combining mass accretion histories of galaxy-size haloes with a chemical
evolution model for individual satellites.  In their model, the agreement with
the observed trends is a consequence of the fact that the stellar halo
originates from a few (relatively massive) satellites accreted early on and
enriched in $\alpha$-elements by type II supernovae. The surviving satellites
are typically accreted much later, have more extended star formation histories
and stellar populations enriched to solar level by both type II and type Ia
supernovae. A more serious problem with the \citeauthor{Searle_and_Zinn_1978}
scenario has been pointed out by \citet{Helmi_etal_2006b} who found a
significant difference between the metal-poor tail of the dSph metallicity
distribution and that of the Galactic halo, demonstrating that the progenitors
of present day dSph's are fundamentally different from the building blocks of
our Galaxy, even at earliest epochs. One possible solution discussed in Helmi
et al. is that the Galactic building blocks formed from the collapse of
high-$\sigma$ density fluctuations in the early Universe, while the present day
satellites would originate from low-$\sigma$ peaks \citep*[see
also][]{Salvadori_Ferrara_Schneider_2008}.

The formation and the evolution of the baryonic component of galaxies is
regulated by a number of non-linear processes operating on vastly different
scales (e.g. shocking and cooling of gas, star formation, feedback by
supernovae and active galactic nuclei, chemical enrichment, and stellar
evolution). Most of these processes are quite poorly understood even when
viewed in isolation. The difficulties grow considerably when one takes into
account the fact that the physical properties of galaxies are determined by a
complex network of actions, back-reactions, and self-regulation between the
above mentioned physical processes.  In recent years, different approaches have
been developed to link the observed properties of luminous galaxies to the dark
matter haloes within which they reside. Among these, semi-analytic models have
developed into a powerful and widely used tool to study galaxy formation in the
framework of the currently standard model for structure formation. In these
models, the evolution of the baryonic component is modelled invoking `simple',
yet physically and observationally motivated `prescriptions'.  These techniques
find their seeds in the pioneering work by \citet{White_Rees_1978}; they have
been laid out in a more detailed form in the early $90$s
\citep*{White_Frenk_1991,Cole_1991,Kauffmann_White_Guiderdoni_1993} and have
been substantially extended and refined in the last years by a number of
different groups. Modern semi-analytic models of galaxy formation take
advantage of high resolution N-body simulations to specify the location and
evolution of dark matter haloes - which are assumed to be the birthplaces of
luminous galaxies
\citep{Kauffmann_etal_1999,Benson_etal_2000,Springel_etal_2001,Hatton_etal_2003}.
Using this `hybrid' approach, it is possible not only to predict observable
physical properties such as luminosities, metallicities, star formation rates,
etc., but also to provide full spatial and kinematical information of model
galaxies thus allowing more accurate and straightforward comparisons with
observational data to be carried out.  In this paper, we use this hybrid
approach to study the formation of the Milky Way galaxy and of its stellar
halo.

The numerical simulations used in our study are described in
Sec~\ref{sec:sims}, while in Sec.~\ref{sec:sam} we give a brief introduction to
the adopted semi-analytic technique, and details of the specific model used in
our study. In Sec.~\ref{sec:prop}, we compare model results to the observed
properties of the Milky Way, and discuss their dependence on a number of model
parameters. In Sec.~\ref{sec:agemet}, we study the age and metallicity
distribution of the spheroid and disk components of our model Galaxy. In
Sec.~\ref{sec:starpart}, we analyse the formation and structure of the stellar
halo. We discuss and summarise our results, and give our conclusions in
Sec.~\ref{sec:discconcl}.

\section{The Simulations}
\label{sec:sims}

In this paper, we use the re-simulations of a `Milky Way' halo (the GA series)
described in \citet{Stoehr_etal_2002} and \citet{Stoehr_etal_2003}. The
underlying cosmological model is a flat $\Lambda$-dominated CDM Universe with
cosmological parameters: $\Omega_{\rm m}=0.3$, $\Omega_{\Lambda}=0.7$, $H_0 =
70\,{\rm km}\,{\rm s}^{-1}\,{\rm Mpc}^{-1}$, $n=1$, and $\sigma_8=0.9$.  The
simulations were generated using the `zoom' technique \citep*{Tormen_etal_1997}
starting from an intermediate-resolution simulation (particle mass $\sim 10^8
M_{\sun}$) of a `typical' region of the Universe.  A `Milky Way' halo was
selected as a relatively isolated halo which suffered its last major merger at
$z > 2$, and with approximately the correct peak rotation velocity.  These
choices were intended to select a candidate for resimulations with properties
that closely match the observed ones: our Galaxy is believed to have had a
quiet merging history, with the last major merging event occurring some 10-12
Gyr ago \citep{Gilmore_Wyse_Norris_2002}, and it is a quite isolated galaxy
with the closest cluster (Virgo) lying at a distance of about 20 Mpc.

It is important to note that haloes with such quiet merging histories are not
uncommon in $\Lambda$CDM cosmologies. Using the Millennium Simulation
\citep{Springel_etal_2005}, we determined the epoch of the last accretion event
corresponding to an increase in mass of the main progenitor of a factor $\sim
0.3$ (i.e. a `major merger') for haloes with present-day mass in the range
$1\,{\rm to}\,3\times10^{12}\,{\rm M}_{\sun}$. We find that about 57 per cent
of these haloes experienced their last `major merger' at redshift larger than
2. The estimated fraction should be considered as an upper limit, as part of
the accretion might occur through diffuse material.  Lower - but still
significant - fractions (about 45 per cent) have been found in a recent
analysis by \citet{Stewart_etal_2007}, and in other independent studies
(Boylan-Kolchin, private communication).

The selected halo was then re-simulated at a series of four progressively
higher resolution simulations using the code {\small GADGET}
\citep*{Springel_Yoshida_White_2001}. The numerical parameters used in the high
resolution regions of these re-simulations are summarised in
Table~\ref{tab:tab1}.

\begin{table}
\caption{Numerical parameters of the re-simulations used in this study. In the 
  table, we give the particle mass $m_{\rm p}$, the number of particles, the
  starting redshift $z_{\rm start}$, and the gravitational softening
  $\epsilon$. All these quantities refer to the particles in the high
  resolution region only.}  

\begin{tabular}{llllll}
  \hline
Name & $m_{\rm p}$ [$h^{-1}$M$_{\odot}$] & ${\rm N}_{\rm HR}$ & $z_{\rm start}$ & $\epsilon$ [$h^{-1}$kpc]\\
\hline
GA0 & $1.677\times10^8$ & 68323    & 70 & 1.4 \\

GA1 & $1.796\times10^7$ & 637966   & 80 & 0.8\\
              
GA2 & $1.925\times10^6$ & 5953033  & 90 & 0.38\\
              
GA3 & $2.063\times10^5$ & 55564205 & 60 & 0.18\\
\hline
\end{tabular}
\label{tab:tab1}
\end{table}

Simulation data were stored in 108 outputs from $z=37.6$ to $z=0$.  These are
approximately logarithmically spaced in time down to $z=2.3$, and approximately
linearly spaced in time thereafter. For each simulation snapshot, we
constructed group catalogues using a standard friends--of--friends (FOF)
algorithm with a linking length of $0.2$ in units of the mean particle
separation. Each group was then decomposed into a set of disjoint substructures
using the algorithm {\small SUBFIND} \citep{Springel_etal_2001}. As in previous
work, we consider to be genuine substructures only those with at least $20$
bound particles which sets the subhalo detection limit to $4.79\times10^9$,
$5.13\times10^8$, $5.5\times10^7$, and $5.89\times10^6\,{\rm M}_{\sun}$ for the
four simulations used in our study.  Substructure catalogues were then used to
construct merger history trees for all self-bound haloes as described in detail
in \citet{Springel_etal_2005} and \citet{DeLucia_Blaizot_2007}. We recall that
the merger tree construction is based on the determination of a unique
descendant for any given halo. In order to determine the appropriate
descendant, for each halo we find all haloes in the following snapshot that
contain its particles, and then count the particles giving higher weight to
those that are more tightly bound in the halo under consideration.  As a
reference, \cite{Springel_etal_2005} gave about three times more weight to the
half most bound particles of a halo of 100 particles.  (The relative weight of
the half most bound particles increases with particle number.) For our
analysis, we are interested in tracing well the cores of the accreted
satellites (see Sec.~\ref{sec:starpart}), and we noted that the tree
construction adopted in previous work lead to occasional premature mergers. In
order to avoid these events, we have slightly increased the weight of the most
bound particles (four times more weight is given to the half most bound
particles of a 100 particle halo). The merger trees constructed as described
above, represent the basic input needed for the semi-analytic model which is
described in the next section.

\begin{figure*}
\bc
\resizebox{7.5cm}{!}{\includegraphics[]{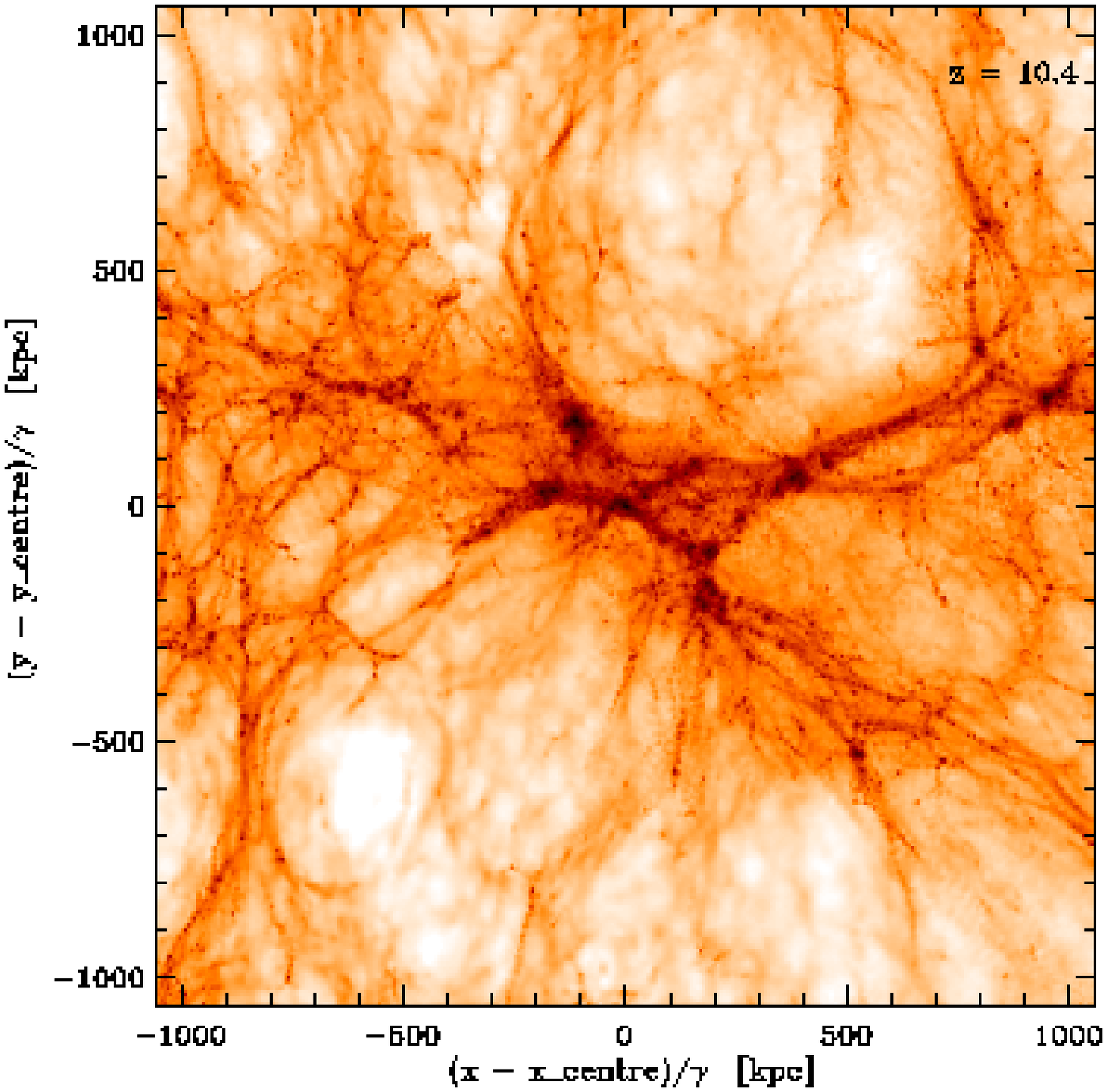}}
\resizebox{7.5cm}{!}{\includegraphics[]{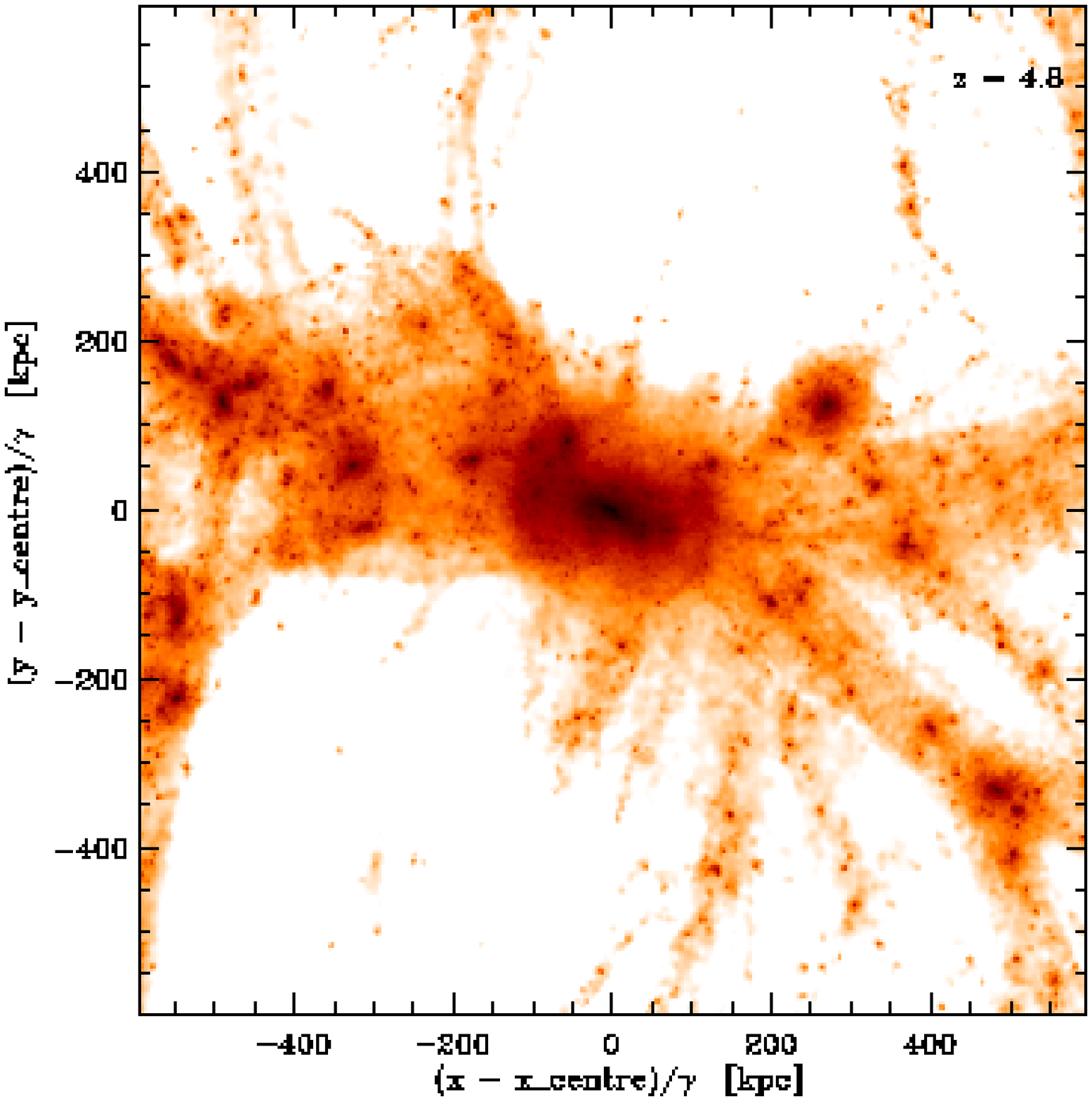}}
\resizebox{7.5cm}{!}{\includegraphics[]{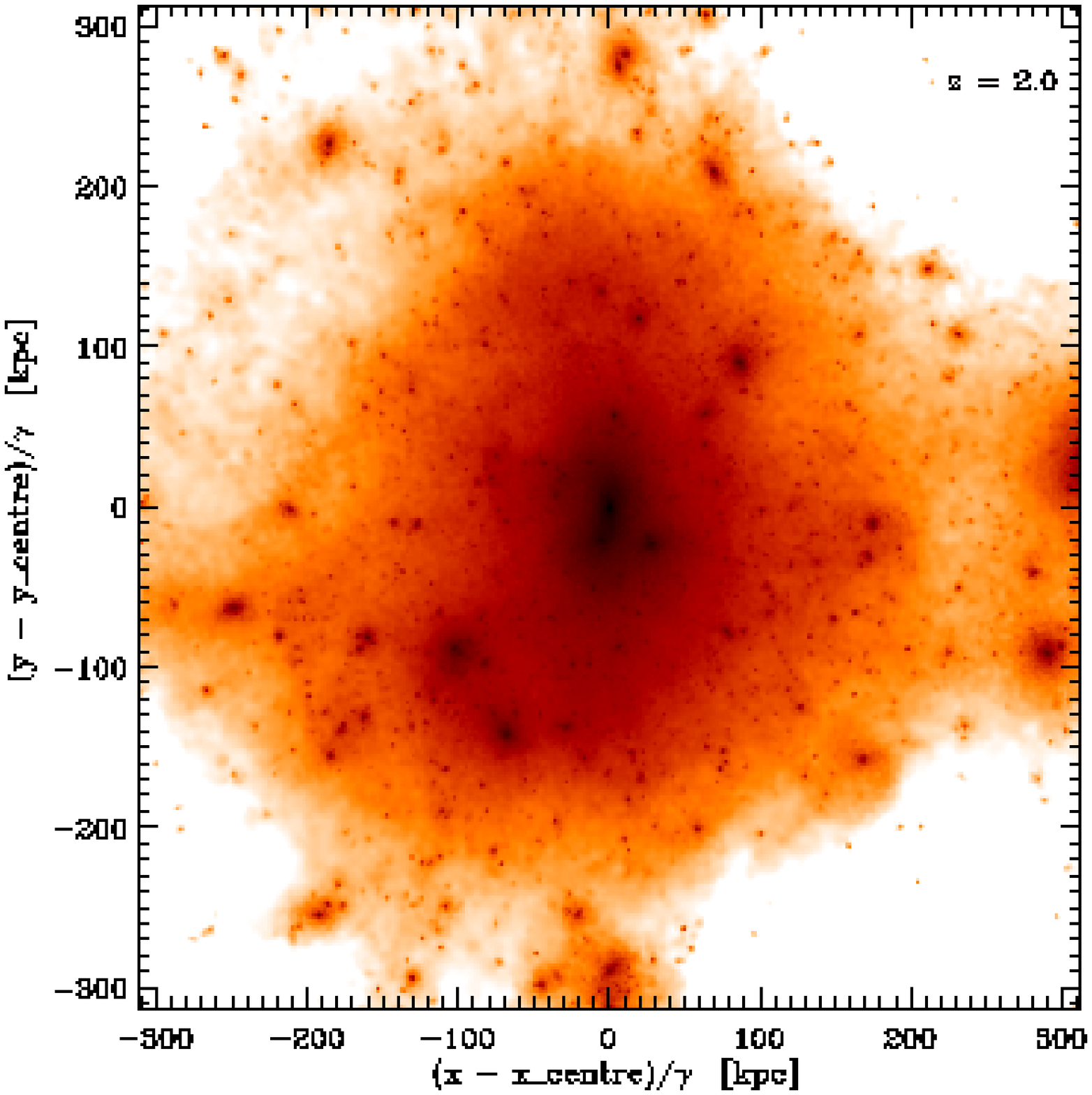}}
\resizebox{7.5cm}{!}{\includegraphics[]{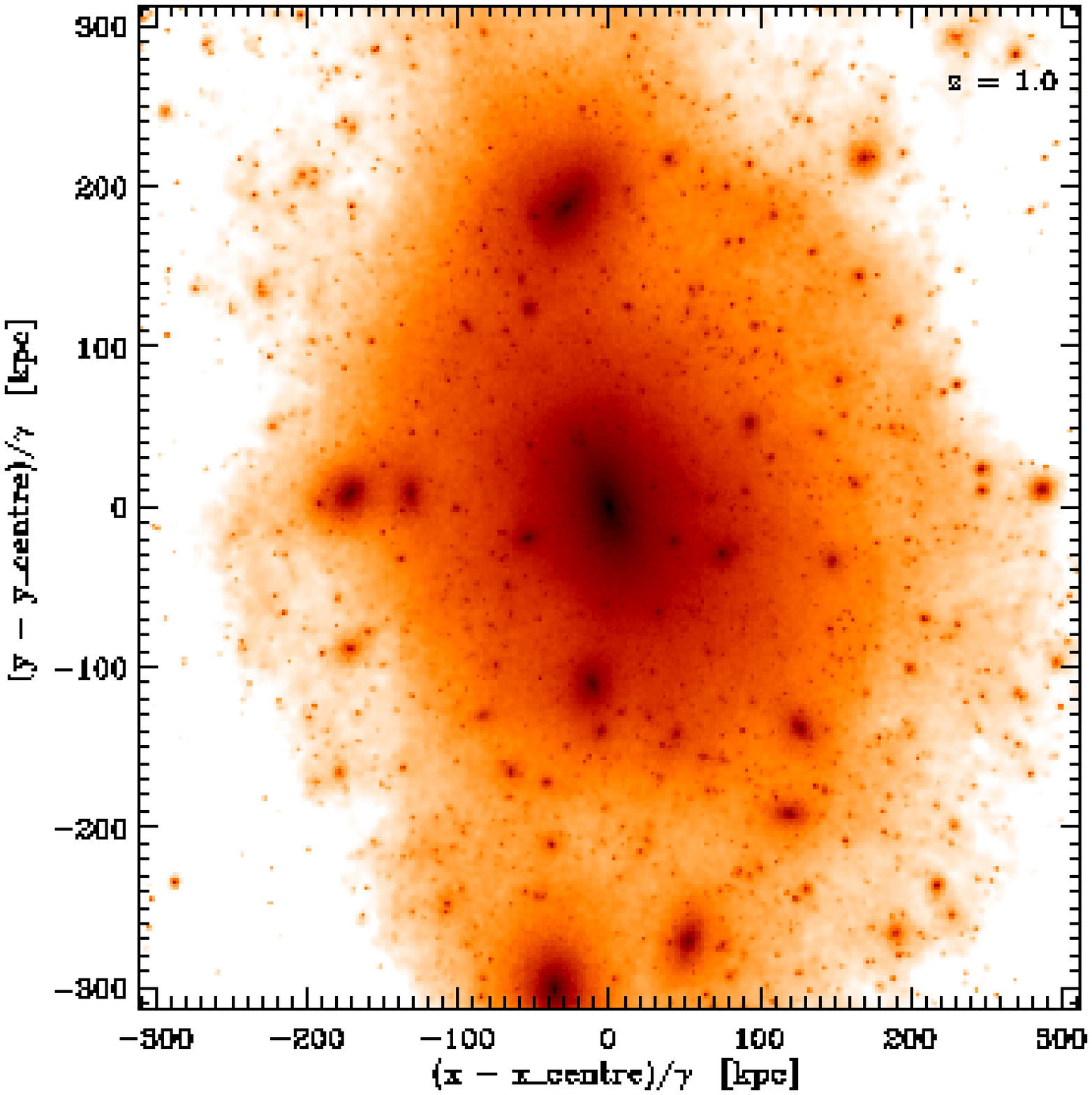}}
\resizebox{7.5cm}{!}{\includegraphics[]{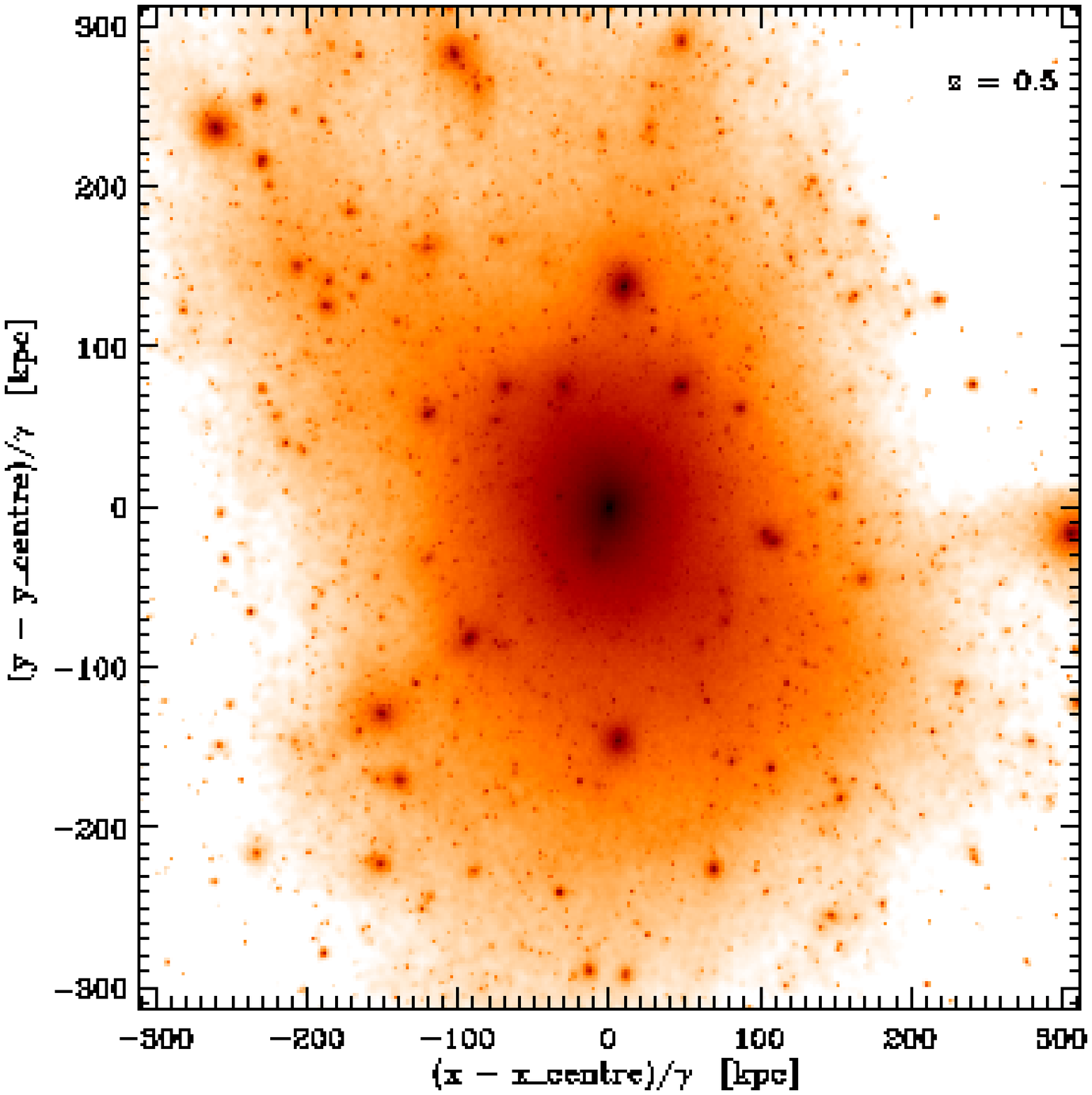}}
\resizebox{7.5cm}{!}{\includegraphics[]{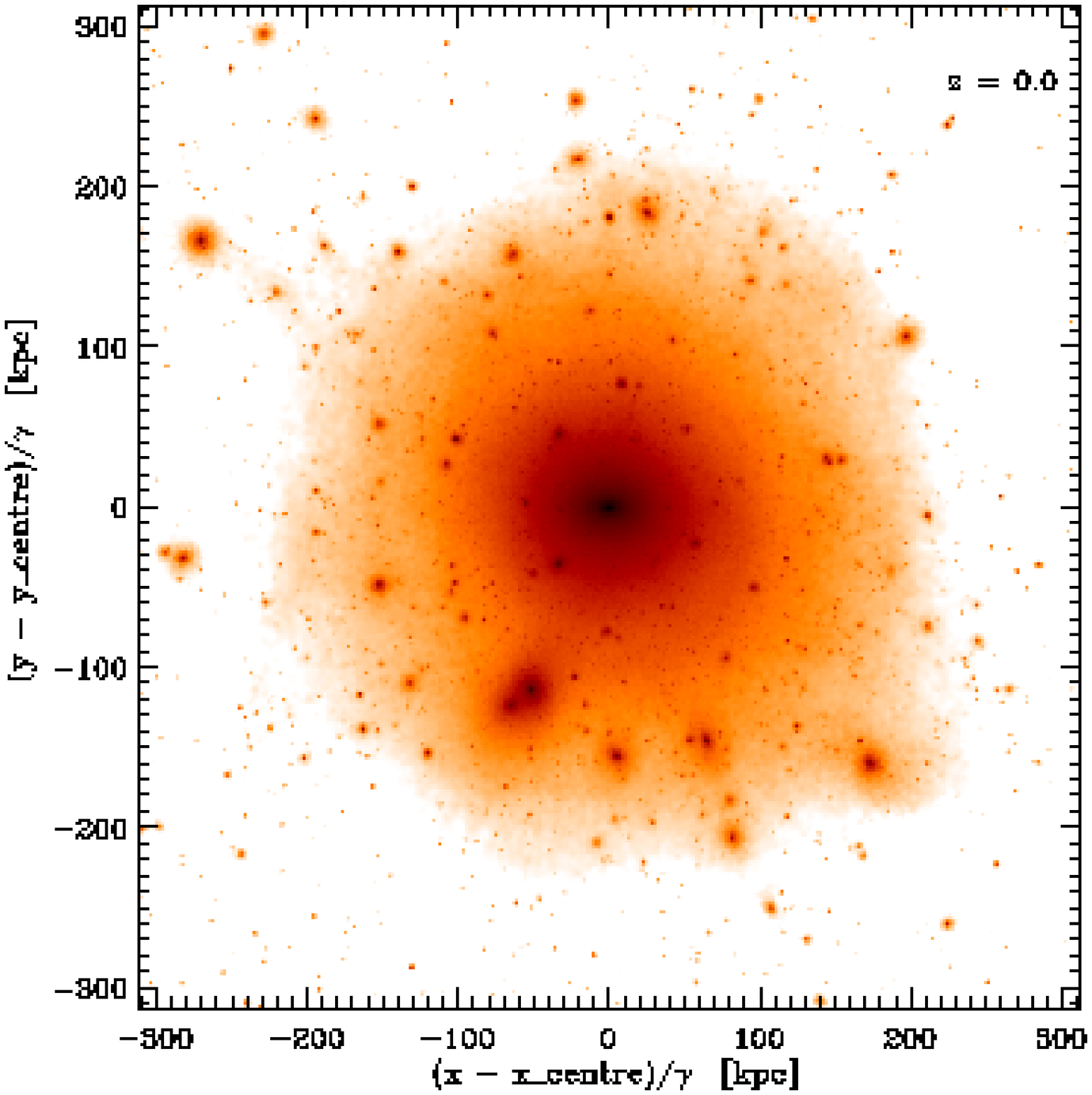}}
\caption{Projected density distribution of the dark matter for the simulation
  GA3, at six different redshifts (from top left to bottom right: 10.4,
  4.8,2.0,1.0,0.5, and 0). The box at $z=0$ is centred on the most bound
  particle of the GA3 halo, while those at higher redshift are centred on the
  most bound particle of the main progenitor of the GA3 halo at the
  corresponding redshift. For all panels, the depth of the box is $500\,{\rm
    kpc}$ comoving. As explained in the text, positions are scaled by a factor
  $\gamma = 1.42$.}
\label{fig:GA3maps}
\ec
\end{figure*}

In Table~\ref{tab:tab2}, we list the masses (${\rm M}_{200}$) and maximum
velocities for the main halo in the four simulations used in this study.
\footnote{${\rm M}_{200}$ is defined here as the mass within a sphere of
  density 200 times the critical density.} Latest observational results give
${\rm V}_{\rm max} \sim 180 - 220 \,{\rm km}\,{\rm s}^{-1}$ and a mass ${\rm
  M}_{\rm MW} \sim 1\times10^{12}\,{\rm M}_{\odot}$
\citep{Battaglia_etal_2005,Smith_etal_2007}.  Our simulated haloes therefore
appear more massive than the Milky-Way halo.  Following
\citet{Helmi_White_and_Springel_2003}, we scale our simulations to a `Milky
Way' halo by adopting a scaling factor in mass ${\rm M}_{200}/ {\rm M}_{\rm
  MW} = \gamma^3 = 2.86$. This implies that we scale down the positions and
velocities by the same factor $\gamma = 1.42$ in all four simulations.

\begin{table}
\caption{${\rm M}_{200}$ and ${\rm V}_{\rm max}$ for the four simulations used
  in this study.} 

\begin{tabular}{llllll}
  \hline
Name & ${\rm M}_{200}$ [${\rm M}_{\odot}$] &  ${\rm V}_{\rm max}$ [${\rm
  km}\,{\rm s}^{-1}$]\\
\hline
GA0 & $3.35\times10^{12}$ & 250.71\\
                        
GA1 & $3.28\times10^{12}$ & 246.60\\
                        
GA2 & $3.21\times10^{12}$ & 247.05\\
                        
GA3 & $2.98\times10^{12}$ & 251.20\\
\hline
\end{tabular}
\label{tab:tab2}
\end{table}

Fig.~\ref{fig:GA3maps} shows the projected density distribution of the dark
matter for the simulation GA3, at six different redshifts. The box at $z=0$ is
centred on the most bound particle of the GA3 halo, while those at higher
redshift are centred on the most bound particle of the main progenitor of the
GA3 halo at the corresponding redshift.

\section{The hybrid model of galaxy formation}
\label{sec:sam}

A complete review of semi-analytic techniques goes beyond the scope of this
paper. For a pedagogical and historical introduction to semi-analytic methods,
we refer the interested reader to the recent review by \citet{Baugh_2006}. In
this section, we briefly summarise how a semi-analytic model is grafted onto
high resolution $N$-body simulations, and give a brief account of those aspects
of the physical model which are relevant for the present study.

The basic assumption is that galaxies form when gas condenses at the centre of
dark matter haloes. Star formation, feedback processes, chemical enrichment,
etc. take then place according to analytical laws which are based on
theoretical and/or observational arguments. Adopting this formalism, it is
possible to express the full process of galaxy evolution through a set of
differential equations that describe the variation in mass as a function of
time of the different galactic components (e.g.  stars, gas, metals), and that
are coupled to the merger history of the dark matter haloes extracted from the
$N$-body simulations \citep*[e.g. see Fig.~1 and Sec.~4.7
in][]{DeLucia_Kauffmann_White_2004}. Given our limited understanding of the
physical processes that regulate galaxy formation and evolution, the equations
describing these processes contain `free' parameters whose value is typically
chosen in order to provide a reasonably good agreement with the observational
data in the local Universe.

The semi-analytic model used in this study builds on the methodology originally
introduced by \citet{Springel_etal_2001} and
\citet{DeLucia_Kauffmann_White_2004}, and has been recently updated to include
a model for the suppression of cooling flows by `radio-mode' AGN feedback as
described in detail in \citet{Croton_etal_2006}. Details about the modelling
adopted for the various physical processes considered can be found in
\citet{Croton_etal_2006} and \citet{DeLucia_Blaizot_2007}.

We recall that our approach follows explicitly dark matter haloes when they are
accreted into larger systems. This allows the dynamics of satellite galaxies
residing in the infalling haloes to be properly followed until their parent
dark matter substructures are completely destroyed by tidal truncation and
stripping \citep{Ghigna_etal_2000,DeLucia_etal_2004,Gao_etal_2004}. When this
happens, the satellite galaxy residing at the centre of the substructure under
consideration is assumed to merge onto the central galaxy after a residual
surviving time that is estimated from the relative orbit of the two
  merging objects, at the time of subhalo disruption:
\begin{displaymath}
  T_{\rm merge} = \,1.17\cdot f_{\rm fudge}\cdot
    \frac{r_{\rm sat}^2\cdot V_{\rm virial}}{\ln\,\Lambda \cdot G \cdot M_{\rm
    sat}} 
\end{displaymath}
In the above equation, $M_{\rm sat}$ is the mass of the substructure at the
last time it is identified, $r_{\rm sat}$ is the distance between the merging
halo and the centre of the structure on which it is accreted, $V_{\rm virial}$
is the virial velocity of the accreting structure, and we approximate the
Coulomb logarithm with ${\rm ln}\,\Lambda = (1+M_{\rm vir}/M_{\rm sat})$. We
therefore allow the satellite galaxy to further sink in through dynamical
friction, even after its dark matter halo has fallen below the resolution limit
of our simulations.  When a halo is accreted onto a larger structure, its mass
is reduced by tidal stripping and its orbit is shrunk by dynamical friction,
until the substructure can no longer be identified as a self-bound overdensity
orbiting the smooth dark matter background of the larger system. This
`disruption' of the dark matter substructure typically occurs at $r_{\rm sat}
\ge\, 0.1 R_{\rm vir}$ which can be much larger than the separation from which
the galaxy merger is expected to happen. We note, finally, that previous work
has shown that this treatment of orphan galaxies is required in order to
reproduce the correlation signal at small scales \citep{Wang_etal_2006,
  Kitzbichler_White_2008}.

Mild variations of the above formula are used in different semi-analytic models
with differences entering mainly the way each model treats the orbital
distribution, ${\rm ln}\,\Lambda$, and $f_{\rm fudge}$ \citep*[see discussion
  in][]{Boylan-Kolchin_Ma_Quataert_2008}. In this work, we have followed
\citet{DeLucia_Blaizot_2007} and used $f_{\rm fudge} = 2$, which is in better
agreement with recent numerical work indicating that the classical dynamical
friction formulation tends to under-estimate the merging times measured from
simulations \citep{Boylan-Kolchin_Ma_Quataert_2008,Jiang_etal_2008}. Results of
these studies, however, are still quantitatively different and additional work
is needed in order to calibrate the correct pre-factor and/or corrections to
apply to the classical formula.

The stellar mass of the satellite galaxies is assumed to be unaffected by the
tidal stripping process that reduces the mass of its parent halo. However, such
tidal effects may well dominate the disruption process of satellites orbiting
in a galaxy halo \citep{penarrubia-benson}.  Therefore, our implementation
could lead to an underestimate of the stellar mass in diffuse form, and to an
overestimate of the number and/or luminosity of satellite galaxies.  We will
see however, that this does not affect our conclusions in any significant way.

As in previous work \citep{Croton_etal_2006,DeLucia_etal_2006}, we assume that
spheroid formation occurs through both mergers and disk instability. In the
case of a `minor' merger, we transfer the stellar mass of the merged galaxy to
the spheroid of the central galaxy. The cold gas of the satellite galaxy is
added to the disk of the central galaxy, and a fraction of the combined cold
gas from both galaxies is turned into stars as a result of the merger. Any
stars that formed during the burst are also added to the disk of the central
galaxy.  The photometric properties of the galaxy are updated accordingly using
the same method described in \citet{DeLucia_Kauffmann_White_2004}. If the mass
ratio of the merging galaxies is larger than $0.3$, we assume that we witness a
`major' merger that gives rise to a more significant starburst and destroys the
disk of the central galaxy completely, leaving a purely spheroidal stellar
remnant.  The galaxy can grow a new disk later on, provided it is fed by an
appreciable cooling flow.  Note that the modelled spheroidal component includes
both the bulge and the stellar halo.

Following \citet*{Mo_Mao_White_1998}, we assume that a stellar disk becomes
unstable when the following condition is verified:
\begin{displaymath}
  \frac{V_{\rm max}}{({\rm G}\,m_{\rm disk}/r_{\rm disk})^{1/2}} \lesssim 1.1
\end{displaymath}
This condition is based on the numerical work by \citet*{Efstathiou_etal_1982}
who used $N$-body simulations to investigate the development of global
instabilities in exponential disks embedded in a variety of haloes. In this
work, we assume that each time a galaxy meets the above criterion, a fixed
fraction ($F_{\rm inst}$) of the stellar mass in the disk is transferred to the
central spheroid component. The instability criterion is applied only to disk
dominated systems (${\rm M}_{\rm spheroid}/{\rm M}_{\rm tot} < 0.1$). This
choice is motivated by numerical studies that showed that a spheroid component
can stabilise the disk against bar formation, giving the galaxy an inner
Lindblad resonance which does not allow swing amplification of waves through
the centre \citep[e.g.][]{Sellwood_1989,Sellwood_and_Moore_1999}.  The
parameter $F_{\rm inst}$ is chosen so as to get a morphological mix that is in
reasonable agreement with the observational measurements for the local
Universe\footnote{The morphological mix resulting from the adopted model has
  been computed for the Millennium Simulation.}, and is assumed to be equal to
0.5. We note that, according to the condition above, the occurrence of disk
instability episodes follows essentially from the way the disk mass builds up
relative to the total mass of the halo, and that the condition ${\rm M}_{\rm
  spheroid}/{\rm M}_{\rm tot} < 0.1$ efficiently suppresses the occurrence of
late instability episodes even in galaxies with modest bulge-to-total ratios.

Admittedly, our modelling of disk instability is very simplified, and gives a
very crude description of the complex phenomenology associated to bar formation
and evolution, which is the subject of current active research \citep[][and
  references
  therein]{Martinez-Valpuesta_etal_2006,Berentzen_etal_2007,Curir_etal_2007}.
We are, for example, neglecting the possibility that bar formation produces an
inflow of gas towards the centre that could fuel starburst/AGN activity, and
that can eventually lead to bar disruption. In addition, present simulations do
not provide clear indications about the fraction of disk mass that gets
re-distributed and how this depends on the halo/galaxy properties. We therefore
consider the results of our simple modelling as just indicative and, in the
following, we also comment on results obtained when the disk instability
channel for bulge formation is switched off.

In this paper, we adopt an improved model to estimate the disk radii. Assuming
conservation of specific angular momentum, we assume that when hot gas cools at
the centre of dark matter haloes, it settles in a rotationally supported disk
with exponential scale-length given by:
\begin{displaymath}
  r_{D} = \frac{\lambda}{\sqrt{2}}R_{200}
\end{displaymath}
where $\lambda$ is the halo spin parameter \citep{Mo_Mao_White_1998}.
Following \citet{Hatton_etal_2003}, we recompute the scale-length at each
time-step by taking the mass-weighted average gas profile of the disk and that
of the new material being accreted. The scale-length of the disk is not altered
after a galaxy becomes a satellite.

As in previous work, we assume that the star formation occurs at a rate given
by:
\begin{displaymath}
  \psi = \alpha_{\rm SF} M_{\rm sf} / t_{\rm dyn}
\end{displaymath}
where $t_{\rm dyn} = r_{\rm disk}/V_{\rm vir}$ is the dynamical time of the
galaxy and we assume that the star forming region ($r_{\rm disk}$) extends to
about $3\times r_{D}$. The parameter $\alpha_{\rm SF}$ regulates the efficiency
of the conversion of gas into stars, and $M_{\rm sf}$ represents the amount of
gas available for star formation which can be expressed as:
\begin{displaymath}
  M_{\rm sf} = 2\pi \int_0^{r_{\rm crit}} \Sigma_{D}(r)rdr 
\end{displaymath}
where $r_{\rm crit}$ is the radius at which the gas surface density drops below
the following critical value \citep{Kennicutt_1989}:
\begin{displaymath}
  \Sigma_{\rm crit} [{\rm M}_{\sun}\,{\rm pc}^{-2}] = 0.59\,V [{\rm km}\,{\rm
  s}^{-1}] / r_{\rm disk}\, [{\rm kpc}].
\end{displaymath}

\noindent
For a complete summary of the model parameters adopted in our fiducial model,
we refer to Table 1 of \citet{Croton_etal_2006}. As explained in
\citet{DeLucia_Blaizot_2007}, the adoption of a Chabrier Initial Mass Function
led to slight modifications of some of these parameters. We refer to these
papers for more details about the parametrisation adopted for the various
physical processes explicitly taken into account in our model, and for a
complete characterisation of model parameters.

\section{Dependency on model parameters and numerical resolution}
\label{sec:prop}

\begin{figure*}
\bc
\hspace{-0.45cm}
\resizebox{18cm}{!}{\includegraphics[]{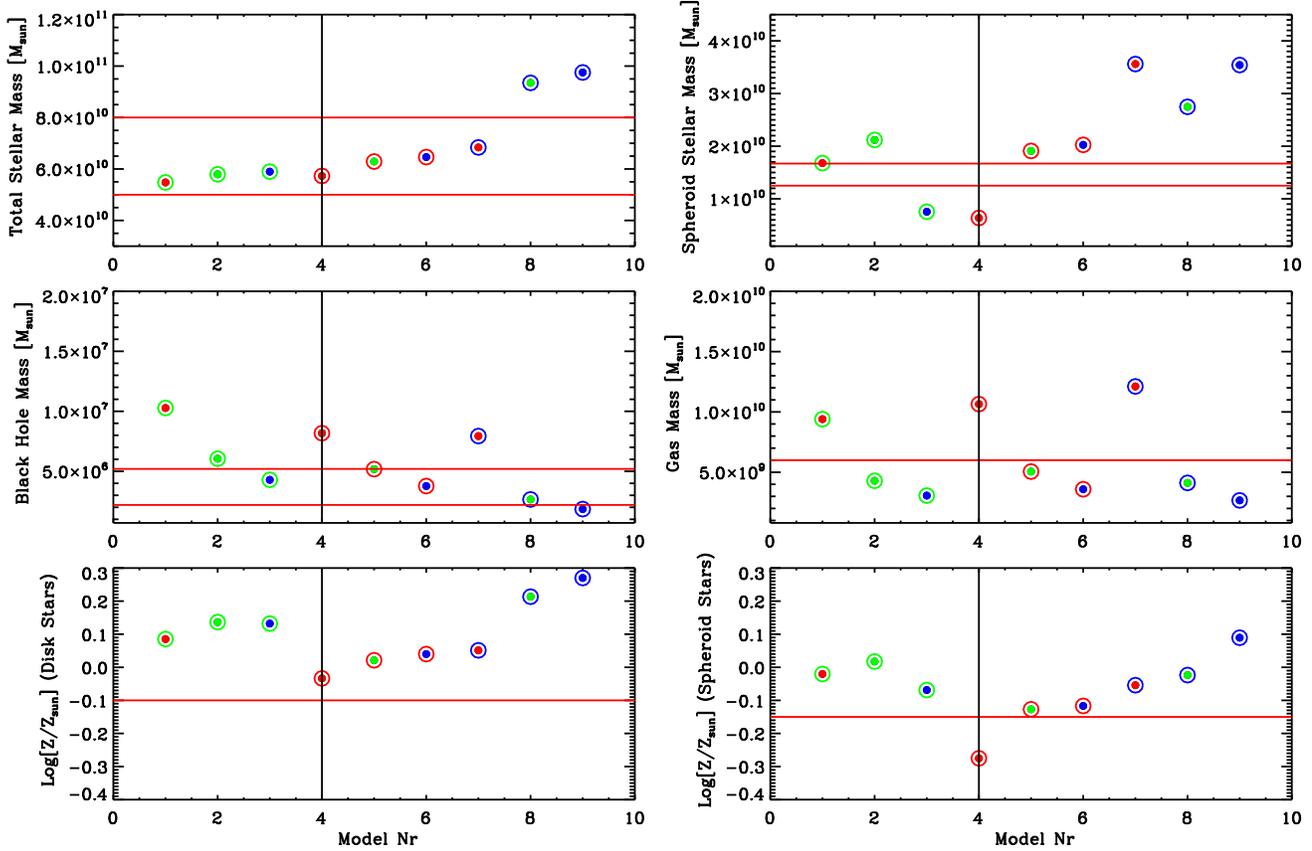}}
\caption{Physical properties of our model Milky Way galaxy from the
  simulation GA3, as a function of different combinations of the star formation
  efficiency ($\alpha_{\rm SF}$), and of the supernovae feedback efficiency
  ($\epsilon_{\rm disk}$). Different colours for filled circles correspond to
  different values for the star formation efficiency parameter: 0.03 (red),
  0.08 (green), 0.13 (blue).  Different colours for open circles correspond to
  different values for the SN feedback efficiency: 1.5 (green), 3.5 (red), 5.5
  (blue).  The vertical black line (corresponding to model number 4) indicates
  our fiducial model.  Red horizontal lines in each panel indicate
  observational estimates. Note that in the bottom right panel we compare the
  metallicity of the spheroid in our models to that of the Galactic bulge.}
\label{fig:grid}
\ec
\end{figure*}

In order to provide an illustration of how model results are affected by
different choices of model parameters, we show in Fig.~\ref{fig:grid} how
different physical properties of our model Milky Way galaxy vary on a limited
model grid where we have altered only the values of the star formation and
supernovae feedback efficiencies.  These correspond to the parameters
``$\alpha_{\rm SF}$'' and ``$\epsilon_{\rm disk}$'' in Table~1 of
\citet{Croton_etal_2006}, and regulate the amount of cold gas that is converted
into stars in a disk dynamical time, and the amount of cold gas reheated by
supernovae explosions.  As explained in the previous section, our fiducial
model uses the same combination of model parameters adopted in
\citet{DeLucia_Blaizot_2007}, and is indicated by the vertical solid line in
Fig.~\ref{fig:grid}.

In the figure, different colours for filled circles correspond to different
values for the star formation efficiency parameter: 0.03 (red), 0.08 (green),
0.13 (blue).  Different colours for open circles correspond to different values
for the SN feedback efficiency: 1.5 (green), 3.5 (red), 5.5 (blue). Red
horizontal lines in each panel indicate the observational measurements. 
Fig.~\ref{fig:grid} shows results from our highest resolution simulation (GA3),
but similar trends are obtained for the lower resolution simulations used in
this study (see below).

Increasing $\alpha_{\rm SF}$ (red, green, blue filled circles) and keeping
$\epsilon_{\rm disk}$ constant (open circles of the same colour) produces an
increase of total mass, and a corresponding decrease in of the amount of cold
gas available. As explained in \citet{DeLucia_Kauffmann_White_2004}, metals are
exchanged between different components proportionally to the exchanged mass so
that the increase in mass is reflected in a parallel increase in the amount of
metals (bottom left panel in Fig.~\ref{fig:grid}). The mass of the spheroid
component (top left panel) varies in a non-monotonic way with increasing star
formation efficiency because disk instability episodes (see previous section)
occur at different times. In our model, black holes grow primarily during
mergers, both by merging with each other and by accretion of cold gas (see Sec.
3.4 of \citealt{Croton_etal_2006}). The black hole mass therefore decreases
with increasing star formation efficiency (middle left panel) as a consequence
of the decrease of the cold gas available.

Increasing $\epsilon_{\rm disk}$ (green, red, blue open circles) and keeping
$\alpha_{\rm SF}$ constant (filled circles of the same colour), the final
amount of gas stays almost constant because gas ejected by satellite galaxies
is rapidly re-incorporated\footnote{Following
  \citet{DeLucia_Kauffmann_White_2004}, we assume that the material that is
  ejected outside the haloes as a consequence of supernovae explosions, can be
  re-incorporated into the hot gas component after a time-scale that is
  proportional to the dynamical time-scale of the halo. Our default model
  adopts the same re-incorporation efficiency as in \citet{Croton_etal_2006}.},
and because the higher level of enrichment of the hot gas component causes a
more efficient cooling. As a consequence, the total stellar mass increases (top
left panel), and the black hole mass decreases slightly (middle left panel).

For our reference model, the total stellar mass is $\sim 6\times10^{10}\,{\rm
  M}_{\sun}$, in very good agreement with the estimated value of $\sim
5-8\times10^{10}\,{\rm M}_{\sun}$. The mass of the spheroidal component is
instead slightly lower than the observed value (assumed to be $\sim 25$ per
cent of the disk stellar mass, or $1.25 - 1.6\times10^{10}\,{\rm M}_{\sun}$ -
\citealt{bissantz}). Studying the proper motion of Galactic Centre stars,
\citet{Shoedel_etal_2002} estimated a central point mass of
$(3.7\pm1.5)\times10^6\,{\rm M}_{\sun}$, which is lower than the value obtained
for our Milky Way in the reference model. This is not surprising given that our
model for the formation of black holes was tuned to match the local relation
between black hole mass and bulge mass \citep{Croton_etal_2006}, and that the
value measured for the Milky Way is offset from this relation. The total mass
of H$_2$ in the Milky Way galaxy is $\sim 1\times10^9\,{\rm M}_{\sun}$ and the
total mass of HI gas is $\sim 5\times10^9\,{\rm M}_{\sun}$ \citep[][and
  references therein]{Blitz_1997}.  Fig.~\ref{fig:grid} shows that our fiducial
model gives a gas mass which is about twice the estimated value, and that this
`problem' could be alleviated by increasing the star formation efficiency. The
stars of the Galactic bulge are found to peak at near solar value ([Fe/H]$\sim
-0.15$), with a relatively wide spread, while the stars in the disk have
[Fe/H]$\sim -0.1$ \citep[][see also
  Sec.~\ref{sec:agemet}]{Freeman_Bland-Hawthorn_2002}.  The corresponding
metallicities for the galaxy in the reference model are not far from the
observed values. Note, however, that the metallicities shown in
Fig.~\ref{fig:grid}, and in the rest of this paper, refer to total
metallicities (and not only to [Fe/H]).  It should be stressed that our model
assumes an instantaneous recycling approximation, i.e. we neglect the delay
between star formation and the recycling of gas and metals from stellar winds
and supernovae.  The model is therefore not able to take into account the
evolution of different element abundances, and in particular it does not
describe well the elements around the iron-peak, which are mainly produced by
supernovae Ia and which typically return their products on longer
timescales. We will come back to this issue later.

Overall, Fig.~\ref{fig:grid} demonstrates that our reference model is in
relatively good agreement with observational measurements (although there are
other combinations of model parameters that also provide a reasonable
agreement).  This is not entirely surprising given that model parameters are
tuned to reproduce the local galaxy luminosity function {\it and} the mass and
luminosity of `Milky-Way' galaxies (see
\citealt{DeLucia_Kauffmann_White_2004}). In previous work, we have shown that
our fiducial model is in quite good agreement with the observed relations
between stellar mass, gas mass, and metallicity
\citep{DeLucia_Kauffmann_White_2004}, the observed luminosity, colour, and
morphology distributions \citep{Croton_etal_2006,DeLucia_etal_2006}, and the
observed two-point correlation functions
\citep{Springel_etal_2005,Wang_etal_2007}. \citet{Kitzbichler_White_2007} have
recently shown that it also agrees reasonably well with the observed galaxy
luminosity and mass functions at higher redshift. In the following, we will
therefore focus on results from our fiducial model. It is worth mentioning at
this point that our default model, which assumes a redshift of reionization
$z\sim 7$ \citep{Croton_etal_2006}, over-predicts the number of observed
satellites. The situation is significantly improved when adopting an earlier
epoch of reionization which does not affect significantly the model Galaxy and
the results presented in this paper. We will present a more detailed analysis
of the satellite population for our model Milky Way in a forthcoming paper.

\begin{figure*}
\bc
\hspace{-0.8cm}
\resizebox{17cm}{!}{\includegraphics[]{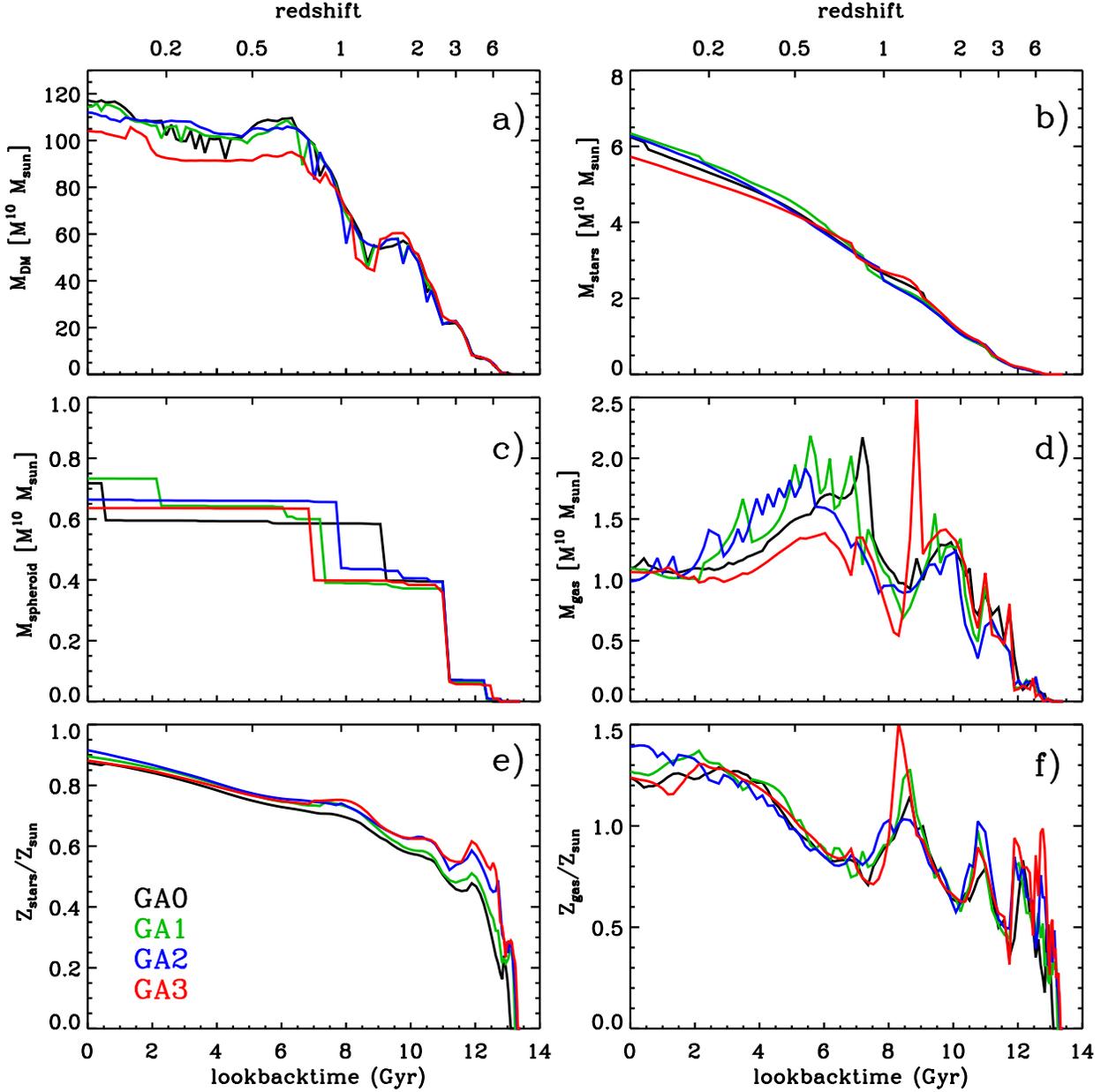}}
\caption{Evolution of the dark matter mass (panel a), total stellar mass (panel
  b), spheroid mass (panel c), cold gas mass (panel d), stellar metallicity
  (panel e), and gas metallicity (panel f) for the model Milky Way galaxies in
  the four simulations used in this study (different colours).}
\label{fig:mgrowth}
\ec
\end{figure*}

Fig.~\ref{fig:mgrowth} shows the evolution of different mass and metallicity
components for the model Milky Way galaxies in the four simulations used in
this study (lines of different colours). The histories shown in panels (a) to
(f) are obtained by tracking the evolution of the `main progenitor', which is
obtained by linking the galaxy at each time-step to the progenitor with the
largest stellar mass. For our model Milky Way, the galaxies merging onto the
main branch have stellar masses that are much smaller than the current mass of
the main progenitor, over most of the galaxy's life-time, explaining the smooth
increase of the stellar mass component.  In these cases, the main progenitor
branch does indeed provide a quite good approximation of the evolution of the
galaxy itself \citep[see discussion in][]{DeLucia_Blaizot_2007}.

Fig.~\ref{fig:mgrowth} shows that approximately half of the final mass in the
dark matter halo is already in place (in the main progenitor) at $z\sim1.2$
(panel a) while about half of the final total (disk + spheroid) stellar mass is
only in place at $z\sim 0.8$ (panel b).  About 20 per cent of this stellar mass
is already in a spheroidal component (panel c). The mass of the spheroidal
component grows in discrete steps as a consequence of our assumption that it
grows during mergers and disk instability events, and approximately half of its
final mass is already in a spheroidal component at $z\sim2.5$. The cold gas
mass content\footnote{Note that in our model, there is no cold gas in the
  spheroidal component. So all gas in panel (d) of Fig.~\ref{fig:mgrowth} is
  associated to the disk.}, in contrast, varies much more gradually (panel d).
Fig.~\ref{fig:mgrowth} also shows that there is a very weak increase of the
stellar metallicity at late times: it only varies from $\sim 0.75$ solar to
$\sim 0.9$ solar over the last 5 Gyr (panel e), while the gas-phase metallicity
varies from $\sim 1$ solar to $\sim 1.3$ solar over the same interval of time.
The more efficient enrichment of the gas component is due to the fact that in
our model all metals produced by new stars are instantaneously returned to the
cold phase, i.e. we are assuming a $100$ per cent mixing efficiency of the
metals with the cold gas already present at the time of star formation.

Interestingly, our model produces consistent evolutions for all four
simulations used in our study, despite a very large increase in numerical
resolution (see Table~\ref{tab:tab1}). This is particularly true for the
stellar mass (panel a) and for the stellar metallicity (panel e). The amounts
of gas (panel d) and metals in the gas phase (panel f) exhibit a much more
noisy behaviour, but the overall evolution is still very similar. We note,
however, that some panels (e.g. panels b, e, and f) do not show convergence of
the results and this is driven by the lack of complete convergence in the
$N$-body simulations (panel a shows a clear difference between GA3 and the
lowest resolution simulations).

\begin{figure}
\bc
\hspace{-0.5cm}
\resizebox{8cm}{!}{\includegraphics[]{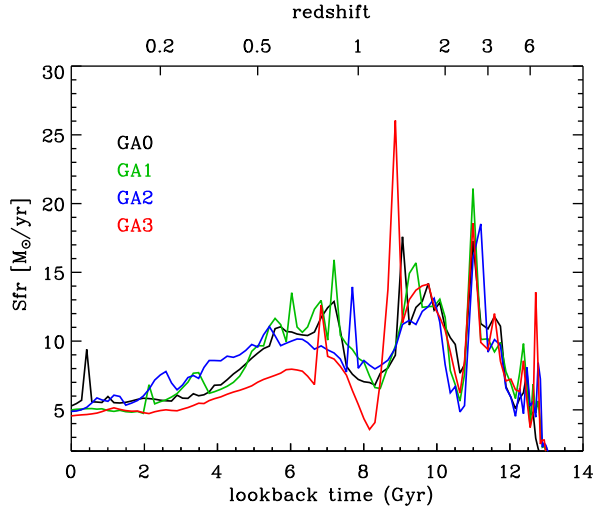}}
\caption{Star formation history of the model Milky Way galaxies from the four
  simulations used in this study (different colours).}
\label{fig:sfr}
\ec
\end{figure}

\begin{figure}
\bc
\hspace{-0.5cm}
\resizebox{8cm}{!}{\includegraphics[]{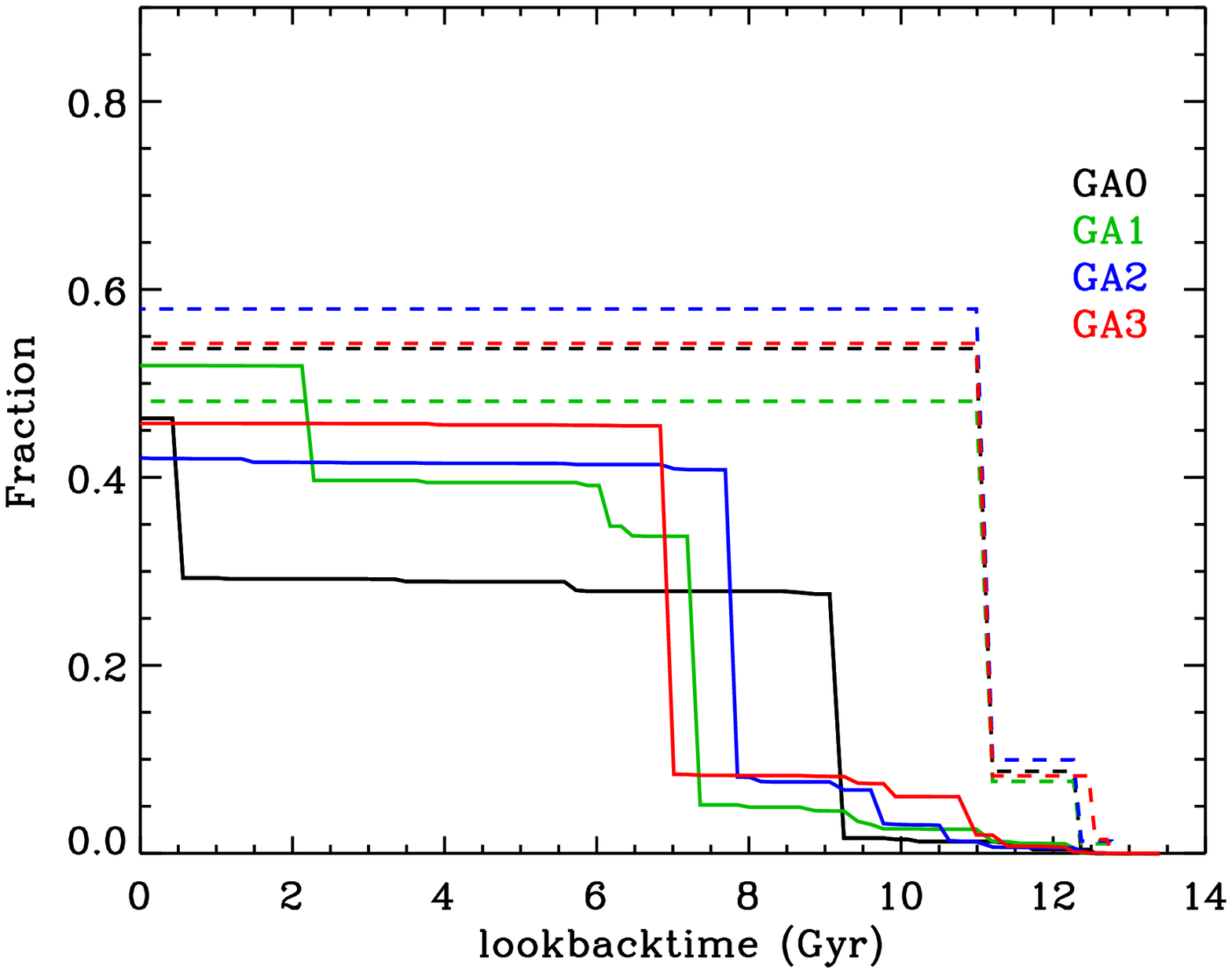}}
\caption{Cumulative fraction of the mass in the spheroidal component coming
  from minor mergers (solid lines) and from disk instability events (dashed
  lines), for the four simulations used in our study (different colours).}
\label{fig:bulge_buildup}
\ec
\end{figure}

Fig.~\ref{fig:sfr} shows the star formation history for the model Milky Way
galaxies from the four simulations used in this study (colour coding as in
Fig.~\ref{fig:mgrowth}). The star formation histories shown in
Fig.~\ref{fig:sfr} exhibit a quite bursty behaviour with several broad periods
of enhanced star formation. In our models, these bursts are triggered by minor
merger events. This intermittent behaviour is in qualitative agreement with the
trend recovered by chromospheric age distributions of late-type dwarfs
\citep[][and references therein]{Rocha-Pinto_etal_2000}. Fig.~\ref{fig:sfr}
also shows that the star formation rate declined slightly over the last $\sim
4$~Gyr and has a current value of $\sim 5\,{\rm M}_{\sun}\,{\rm yr}^{-1}$.
This trend is also in qualitative agreement with observational measurements
\citep*[e.g.][]{Hernandez_Valls-Gabaud_Gilmore_2000,Bertelli_Nasi_2001}, but
the level of star formation activity is significantly higher than the $\lesssim
1\,{\rm M}_{\sun}\,{\rm yr}^{-1}$ estimated for the mean star formation rate in
the Milky Way disc over the last few Gyrs \citep{just-jahreiss}. Although the
general behaviour is very similar for all four simulations used in this study,
there is again no perfect convergence of the results, reflecting the behaviour
of the available cold gas (panel d in Fig.~\ref{fig:mgrowth}).

For our default model and for the four simulations used in this study, the
spheroid mass varies between $6.4\times10^{9}\,{\rm M}_{\odot}$ and
$7.3\times10^{9}\,{\rm M}_{\odot}$ which is slightly lower than but still
compatible with the observational estimate. As explained in the previous
section, our model includes different channels for the formation of a
spheroidal component: mergers (both minor and major) and disk instability.
Fig.~\ref{fig:mgrowth} (panel c) shows how the total mass in the spheroidal
component grows as a function of redshift. In Fig.~\ref{fig:bulge_buildup}, we
show the cumulative fraction of the mass in the spheroidal component coming
from minor mergers (solid coloured lines), and disk instability (dashed lines).
Our model Milky Way galaxies do not experience any major merger during their
life-times. In all four simulations used in our study, there are two episodes
of disk instability, the first $\sim 12$ Gyr ago and the second $\sim 1$ Gyr
later.  Disk instability contributes in the range of 48 (GA1) to 58 (GA2) per
cent of the final stellar mass in the spheroidal component.  Minor mergers
contribute to the remaining stellar mass in the spheroidal component, and all
occur at later times with respect to the disk instability episodes. As
explained in Sec.~\ref{sec:sam}, the stars formed during the bursts
accompanying minor merger events, are added to the disk component. As a
consequence, no spheroid star is formed {\it in situ} in our models.

As noted in Sec.~\ref{sec:sam}, our model for disk instability is quite
simplified and, given the uncertainties involved, the fraction of mass
contributed by this channel should be considered as indicative. The results
shown in Fig.~\ref{fig:bulge_buildup} and Fig.~\ref{fig:grid} indicate that our
default model would not produce a spheroid that is as massive as observed if
the disk instability channel is switched off. On the other hand, it is now
fairly well established that the Milky Way is a barred galaxy \citep[][and
references therein]{Gerhard_2002}. Since global disk instabilities are commonly
considered the main mechanism for the formation of a bar, the presence of a bar
in the Milky Way is a direct indication that it has experienced episodes of
disk instability during its life-time.

\section{Metallicity and age distributions}
\label{sec:agemet}

As mentioned in Sec.~\ref{sec:intro}, accurate measurements of ages,
metallicity and kinematics have been collected for a large number of individual
stars of our Galaxy, and a large amount of new observational measurements are
expected in the near future.  The available information is nicely summarised in
Fig.~2 of \citet{Freeman_Bland-Hawthorn_2002}.

\begin{figure*}
\bc
\hspace{-0.8cm}
\resizebox{17cm}{!}{\includegraphics[]{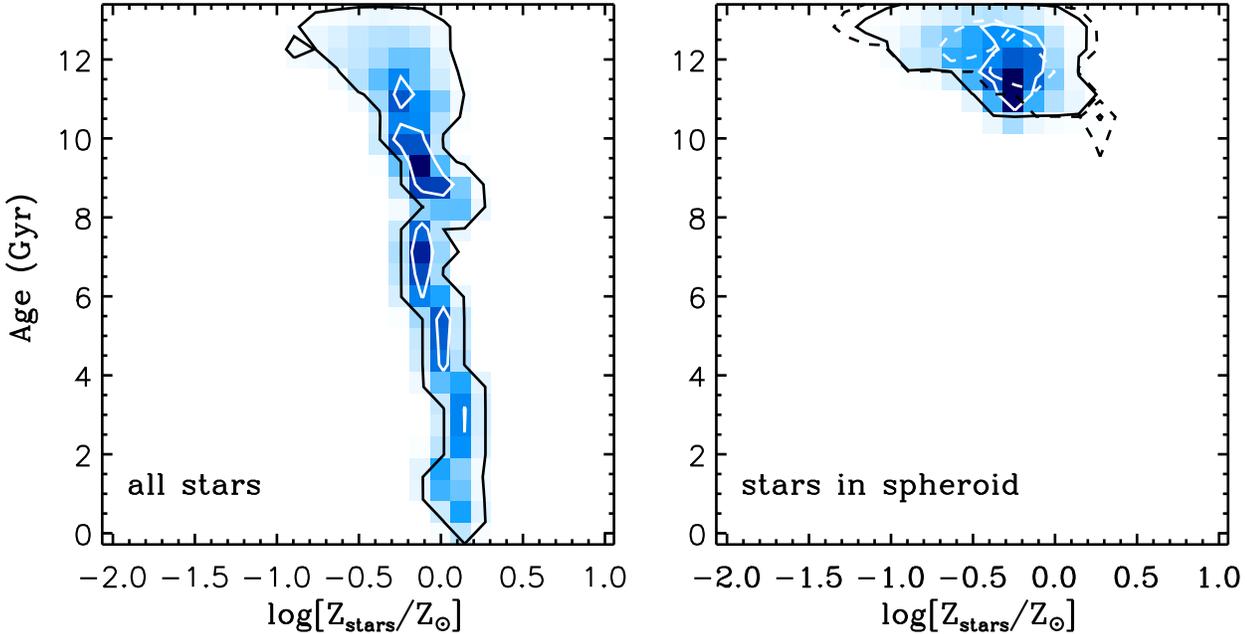}}
\caption{Age-metallicity distribution for all (left panel) and spheroid (right
  panel) stars of the model Milky Way from the highest resolution simulation
  used in this study (GA3). Black and white contour lines show regions that
  include 99.6 and 60 per cent of the stars used to build the corresponding
  map.}
\label{fig:agemetmaps}
\ec
\end{figure*}

The available data are indicative of a significant population of stars in the
local disk with ages $\lesssim 8$~Gyr, and metallicity distribution (mostly
from G/K dwarfs) that peaks at [Fe/H]$ \sim -0.1$. From an analysis of the main
sequence turn-off stars in the Hipparcos dataset,
\citet*{Binney_Dehnen_Bertelli_2000} obtained a best fit age for the oldest
disk stars of $\gtrsim 11$~Gyr. The properties of the thick disk are relatively
well known only within a few kpc from the Sun. In this region, typical thick
disk stars have ages as old as $\sim 12$~Gyr and intermediate metallicities
([Fe/H]$\sim -0.6$). For the Galactic bulge, deep HST and ISO colour-magnitude
diagrams suggest a dominant old ($\gtrsim 10$ Gyr) population
\citep{Feltzing_and_Gilmore_2000,vanLoon_etal_2003}. The ISO data also suggest
the presence of a small intermediate-age component which is traced by OH/IR
stars \citep{Sevenster_1999}, but the interpretation is complicated by thin
disk contamination. Bulge stars have a metallicity distribution that peaks at
[Fe/H]$\sim -0.15$ dex, with a broad range and a tail to low abundances
(\citealt*{McWilliam_and_Rich_1994}, \citealt{Zoccali_etal_2003},
\citealt*{Fulbright_McWilliam_Rich_2006}).

\begin{figure*}
\bc
\hspace{-0.8cm}
\resizebox{17cm}{!}{\includegraphics[]{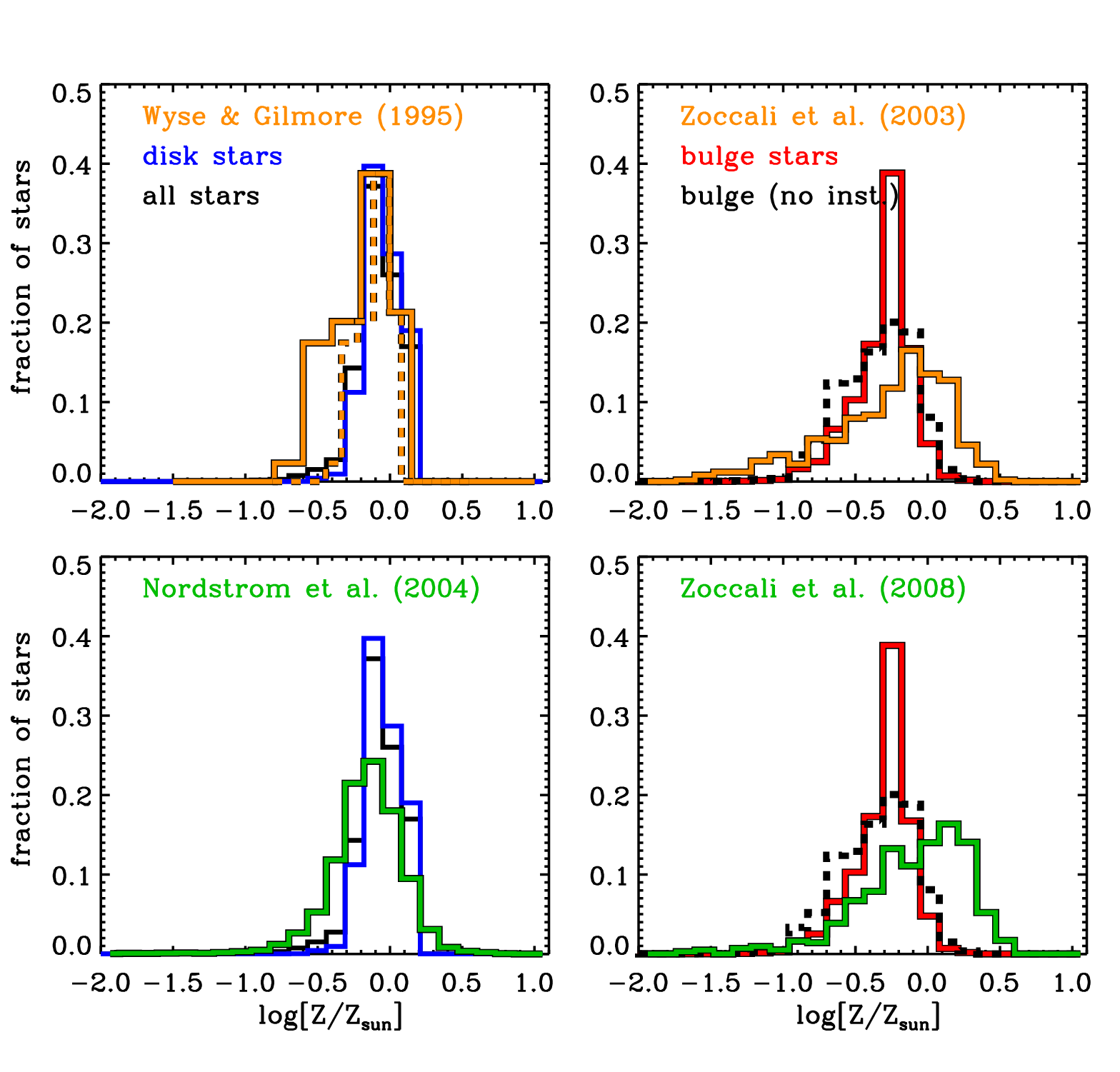}}
\caption{Metallicity distribution for stars in the disk (blue histograms in the
  left panels) and spheroid (red histograms in the right panels) of the model
  Milky Way from the highest resolution simulation in this study (GA3). The
  solid black histograms in the left panels show the metallicity distribution
  for all stars in the model galaxy, while the dashed black histograms in the
  right panels show the metallicity distribution of stars in the spheroidal
  component for our fiducial model if spheroid growth through disk instability
  is suppressed. Solid orange and green histograms show observational
  measurements by \citet[][top left panel]{Wyse_Gilmore_1995}, \citet[][top
  right panel]{Zoccali_etal_2003}, \citet[][bottom left
  panel]{Nordstrom_etal_2004}, and \citet[][bottom right
  panel]{Zoccali_etal_2008}. The dashed orange histogram in the top left panel
  has been obtained converting the [Fe/H] scale of the original distribution by
  \citeauthor{Wyse_Gilmore_1995} into an [O/H] scale by using the observed
  [O/H]-[Fe/H] relation for thin disk stars by
  \citet*{Bensby_Feltzing_Lundstrom_2004}.}
\label{fig:metdistr}
\ec
\end{figure*}

Fig.~\ref{fig:agemetmaps} shows the age and metallicity distribution of all
stars (left panel) and of the stars in the spheroidal component (right panel)
of our model Milky Way galaxy from the highest resolution simulation used in
this study (results for lower resolution simulations are very similar). In each
panel, black and white contours enclose the regions containing 99.6, and 60 per
cent of the stars.  On the right panel, solid contours refer to our fiducial
model where stars can be transferred from the stellar disk component to the
spheroid through disk instability. Dashed lines refer to a model with the same
combination of parameters but with the disk instability channel switched off.
Fig.~\ref{fig:agemetmaps} shows that the Galaxy contains stars of all ages -
which reflects the prolonged star formation history shown in panel (g) of
Fig.~\ref{fig:mgrowth} - and covering a relatively limited range of
metallicities centred around solar.  The stars in the spheroidal component all
have very old ages ($\gtrsim 11$~Gyr), their metallicity distribution peaks at
sub-solar value and exhibit a pronounced tail of low metallicity stars. If
spheroid growth through disk instability is switched off, the stars in the
spheroidal component have a broader tail towards low and high abundances.  The
left panel of Fig.~\ref{fig:agemetmaps} shows a very shallow age-metallicity
relation with a broader tail towards lower metallicities for older stars, in
qualitative agreement with observational measurements
(\citealt{Nordstrom_etal_2004}, see also
\citealt*{Holmberg_Nordstrom_Andersen_2007}). We caution the reader that the
results shown in Fig.~\ref{fig:agemetmaps} represent a global average while
observed samples represent a `local sample'. In addition, the model
age-metallicity distributions are not convolved with typical observational
errors which naturally tend to broaden the distributions.

The metallicity distributions (i.e. the projections along the y-axis of the
maps shown in Fig.~\ref{fig:agemetmaps}) of the stars in the disk and spheroid
of our model galaxies are shown in Fig.~\ref{fig:metdistr}. The left panels
show the metallicity distribution of all stars (black histograms) and of the
stars in the disk (blue histograms) compared to the observational measurements
by \citet{Wyse_Gilmore_1995} (orange histogram in the top panel) and
\citet{Nordstrom_etal_2004} (green histogram in the bottom panel). The sample
of \citet{Wyse_Gilmore_1995} is a volume sample of long-lived thin disk G
stars, while that of \citet{Nordstrom_etal_2004} contains a large ($\sim
14,000$) number of F/G dwarfs in the Solar neighbourhood (located at distances
$< 100$~pc). In both these studies, metallicities were derived using a
photometric calibration based on a relationship between [Fe/H] and Str\"omgren
colours \citep[e.g.][]{schuster-nissen}.

Fig.~\ref{fig:metdistr} thus shows that the metallicity distribution of disk
stars in our model Milky Way peaks at approximately the same value as observed,
but that it exhibits a deficiency of low metallicity stars. In comparing the
model and observed metallicity distributions, however, two factors should be
taken into account: (1) the observational metallicity measurements have some
uncertainties ($\sim 0.2$~dex) which tend to broaden the true underlying
distribution; (2) the metallicity estimate used is an indicator of the {\it
  iron} abundance, which is mainly produced by supernovae Ia and therefore not
well described by the instantaneous recycling approximation adopted in this
study. In order to estimate the importance of this second caveat we have
converted the measured [Fe/H] into [O/H] using a linear relation, which is
obtained by fitting data for thin disk stars from
\citet*{Bensby_Feltzing_Lundstrom_2004}. The result of this conversion is shown
by the dashed orange histogram in the top panel of Fig.~\ref{fig:metdistr} for
the original measurements by \citeauthor{Wyse_Gilmore_1995}. The observed [O/H]
metallicity distribution is now much closer to the modelled log[Z/Z$_\odot$]
distribution.

The right panels in Fig.~\ref{fig:metdistr} show the metallicity distribution
of the spheroid stars in our fiducial model (red histogram) and in a model
where the disk instability channel is switched off (dashed black histograms).
In what follows, it may be useful to associate the stars originating in the
disk instability to the bulge, and the rest of the spheroid stars to the
stellar halo. Model results are compared to observational measurements by
\citet{Zoccali_etal_2003} (orange histogram in the top panel) and recent
results by \citet{Lecureur_etal_2007} and \citet{Zoccali_etal_2008} based on a
sample of about 400 K-giants in two Galactic bulge windows, all observed with
high dispersion spectra (${\rm R} > 20000$) with GIRAFFE on VLT (green
histogram in the bottom panel). These panels show that the metallicity
distribution of the model spheroid peaks at lower value than observed, and that
it contains a smaller fraction of high metallicity stars.  Comparison between
the solid red histogram and the black dashed histogram shows that the disk
instability is responsible for the pronounced peak around
Log[Z/Z$_{\sun}$]$\sim -0.25$ due to the transfer of a large fraction of disk
stars into the spheroidal component, and which may be associated to the
Galactic bulge component (Fig.~\ref{fig:mgrowth} and
Fig.~\ref{fig:bulge_buildup} show that the mean metallicity of the stellar
component is $\sim 0.6\,{\rm Z}_{\sun}$ at the time of the major episode of
disk instability). The same caveat discussed above applies to the metallicity
distribution of the spheroidal components. We note that a conversion from
[Fe/H] to [O/H] of the observed metallicity scale would bring most of the
observed bulge stars to [O/H]$\gtrsim -0.2$ suggesting that our model spheroid
is significantly less metal rich than the observed Galactic bulge. However, we
note that \citet{Fulbright_McWilliam_Rich_2006} find a slightly more metal-poor
distribution for the `Baade's Window', compared to the distribution found for
the same field by \citet[][plotted in the bottom right panel of
  Fig.~\ref{fig:metdistr}]{Zoccali_etal_2008}. In addition, an outer bulge
field (located at $b=-12\deg$) from \citet{Zoccali_etal_2008} peaks at lower
metallicity compared to the metallicity distribution from the other two fields
presented in the same study. This suggests that the uncertainties (as well as
field to field variations) in the metallicity distribution of the bulge may
alleviate the disagreement with our model results.

The metallicity distribution of the remainder spheroid stars (i.e. those
originating in minor mergers) is relatively broad, and more metal-rich than
observed for the Galactic stellar halo.

\section{The stellar halo}
\label{sec:starpart}

\begin{figure*}
\bc
\hspace{-0.6cm}
\resizebox{15cm}{!}{\includegraphics[angle=90]{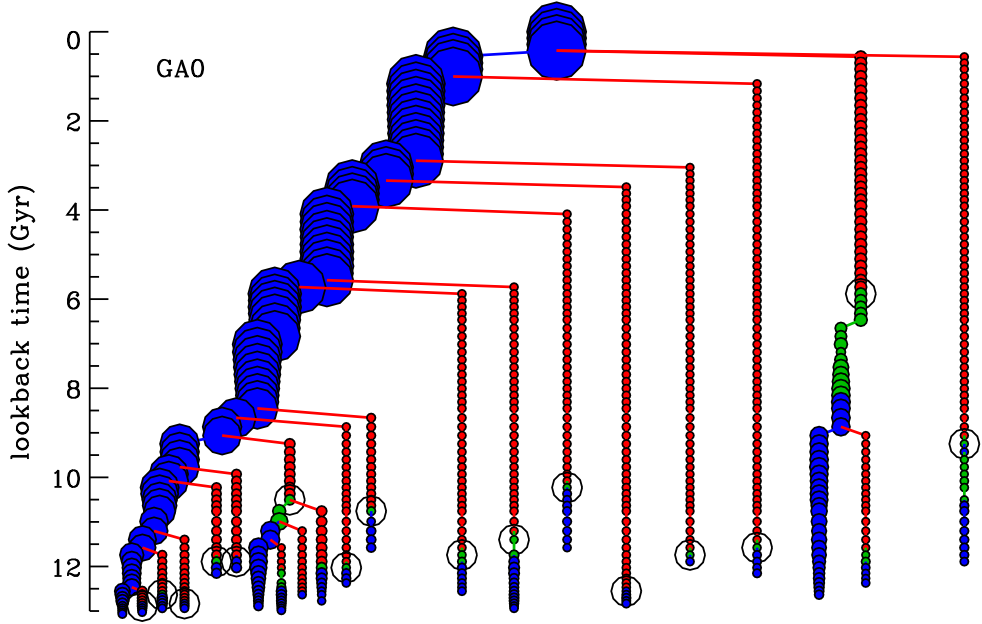}}
\caption{Full merger tree of our model Milky Way galaxy from the lowest
  resolution simulation used in this study (GA0). The area of the symbols scale
  with the number of particles of the associated halo, and their colour
  indicates different galaxy types: blue for galaxies sitting at the centre of
  the main halo (those on which cooling is allowed), green for satellites
  galaxies associated to distinct dark matter substructures, and red for
  `orphan' galaxies (see text). The black circles mark the galaxies accreted
  onto the main branch of the model Galaxy, traced back until the last time
  they are central galaxies.}
\label{fig:ga0tree}
\ec
\end{figure*}

As mentioned in Sec.~\ref{sec:intro}, the stellar halo of our Galaxy has a mass
that is less than about 1 per cent of the total stellar mass (i.e. about
$10^9\,{\rm M}_{\odot}$). Most of its stars are very old (probably older than
12 Gyr) and metal-poor, with enhanced values of the elemental abundance ratio
[$\alpha$/Fe].  The currently accepted scenario is that the stellar halo formed
(at least in part) from stars stripped from satellite galaxies that were
accreted by the Galaxy over its lifetime (the \citeauthor{Searle_and_Zinn_1978}
scenario discussed in Sec.~\ref{sec:intro}). In this section, we study the
formation and structural properties of this galactic component, in the
framework of the model discussed above.

Our working hypothesis is that the stellar halo built up from the cores of the
satellite galaxies that merged with the Milky Way (or better its main
progenitor - see below) over its lifetime.\footnote{Note that some fraction of
  these galaxies may also contribute to the bulge component, particularly those
  with the highest density.} In order to identify the stars that end up in the
stellar halo, we construct the full merger tree of our model Milky Way galaxy,
and identify those galaxies that merge onto the `main branch', i.e. the branch
that is obtained by connecting the system to its most massive progenitor at
each node of the tree (see Sec.~4.1 of \citealt{DeLucia_Blaizot_2007}). We then
trace back each of these galaxies until they are for the last time central
galaxies, and save the information (identification numbers and positions) of a
fixed fraction of the most bound particles of their parent haloes at this time.
We refer to these as `star particles', and `tag' them with the stellar
metallicity of the galaxies residing at their centres.

The procedure outlined above is illustrated in Fig.~\ref{fig:ga0tree} which
shows the full merger tree of our model Milky-Way galaxy from the lowest
resolution simulation used in our study (GA0). The Milky-Way galaxy is shown at
the top of the plot, and all its progenitors and their histories are shown
downward going back in time. In this figure, the area of the symbols scales
with the number of particles in the associated halo, while different colours
are used for different `types' of galaxies. Central galaxies (the only ones on
which cooling is allowed) are shown in blue. Green symbols indicate galaxies
associated to a distinct dark matter substructure, and red symbols indicate
`orphan' galaxies, i.e. galaxies whose parent halo has been reduced below the
resolution limit of the simulation.  For visualisation purposes, we have kept
constant the mass of the dark matter substructure when a galaxy becomes orphan.
The black circles in Fig.~\ref{fig:ga0tree} mark the galaxies accreted onto the
main branch of the model Galaxy (the leftmost branch in the figure), traced
back until the last time they are central galaxies. In the following, we refer
to this time as the {\it time of accretion}. Fig.~\ref{fig:ga0tree} shows that
the star particles that make up the stellar halo in GA0 are old: the majority
of the satellites building up the stellar halo were `accreted' at lookback
times larger than $\sim 9$~Gyr (although about half of them `merge' onto the
main branch much later). Only one (relatively massive) satellite was accreted
at lookback time $\sim 8$~Gyr, and it merges onto the main branch only at
lookback time $\sim 1$~Gyr. For the simulations GA2 and GA3, all accretions
occur at lookback times larger than $\sim 11$~Gyr. As we tag our star particles
using the stellar populations of the central galaxies at the time of accretion,
all stars in the stellar halo of these model galaxies have ages larger than 11
Gyr.

Fig.~\ref{fig:ga0tree} illustrates that once dark matter haloes are accreted
onto a larger system, they survive as distinct dark matter substructures for a
relatively short time \citep[e.g][]{DeLucia_etal_2004}, but the galaxies
residing at their centre merge onto the main progenitor of the Milky Way galaxy
much later. As discussed in Sec.~\ref{sec:sam}, our model galaxies are not
affected by the tidal stripping and truncation that efficiently reduces the
mass of the dark matter substructures
\citep{Ghigna_etal_2000,DeLucia_etal_2004,Gao_etal_2004}. If this effect is
important, the survival time-scales of orphan galaxies, as well as the stellar
mass of survived satellites are likely to be overestimated. Work is underway to
calibrate stellar tidal stripping using hydrodynamic simulations.

\begin{figure}
\bc
\hspace{-0.6cm}
\resizebox{8cm}{!}{\includegraphics[]{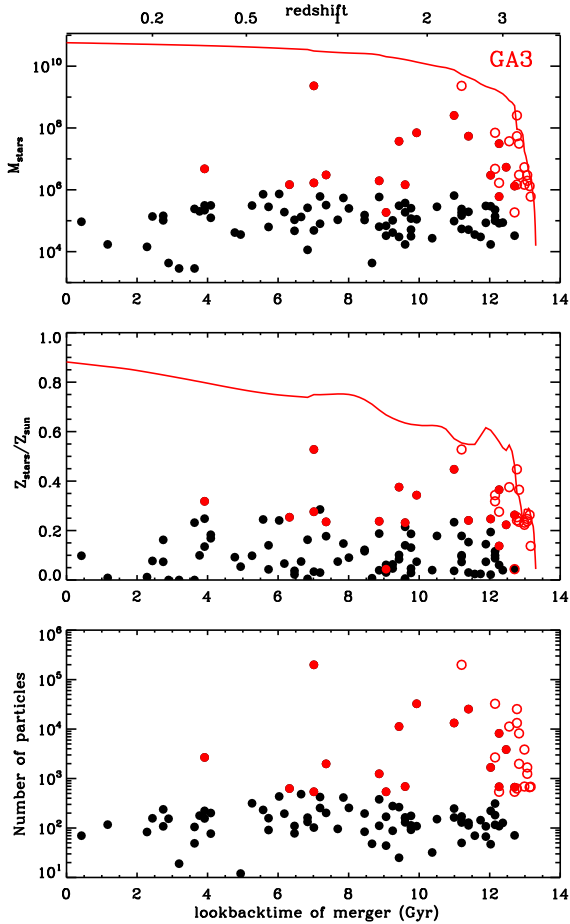}}
\caption{Stellar masses (top panel) and metallicities (middle panel)
  for all galaxies accreted onto the main branch for the simulation GA3, as a
  function of the lookback time of the galaxy's merger. The solid red lines in
  the top and middle panels show the evolution of the stellar mass and
  metallicity in the main progenitor of the Milky Way. The bottom panel shows
  the number of particles associated to the dark matter halo at the time of
  accretion, again as a function of the merging time of the galaxy that is
  located at its centre. Red symbols indicate objects associated to
  substructures with more than 500 particles. Red open symbols correspond to
  red filled circles but are plotted as a function of the time of accretion.}
\label{fig:accretedGA3}
\ec
\end{figure}

Fig.~\ref{fig:accretedGA3} shows the stellar masses (top panel) and
metallicities (middle panel) of all galaxies accreted onto the main branch for
the highest resolution simulation used in our study (GA3), as a function of the
lookback time of the galaxy's merger. Note that stellar masses and
metallicities correspond to those at the time of accretion.  The solid red
lines in these panels show the evolution of the stellar mass and of the stellar
metallicity in the main progenitor of the Milky Way galaxy (as in
Fig.~\ref{fig:mgrowth}). The bottom panel of Fig.~\ref{fig:accretedGA3} shows
the number of particles associated to the dark matter haloes before accretion,
again as a function of the merging time of the galaxies that reside at their
centre. Red symbols indicate objects belonging to haloes with more than 500
bound particles before accretion. Note that for these objects we have indicated
both the time of accretion (open symbols) and the merger time (solid symbols;
given by the dynamical friction timescale as described above).  The figure
shows that most of the galaxies that merge onto the main branch have stellar
masses and metallicities that are much smaller than the current mass of the
main progenitor over most of the galaxy's life-time. Most of the accreted
galaxies lie in quite small haloes and only a handful of them are attached to
relatively more massive systems.  These are the galaxies that contribute most
to the build-up of the stellar halo. The red symbols in the bottom panel of
Fig.~\ref{fig:accretedGA3} show that most haloes with more than $\sim 500$
particles were all disrupted more than $\sim 6$~Gyr ago. These haloes contain
the few galaxies with stellar mass larger than $10^6\,{\rm M}_{\odot}$ which
merge onto the main branch over the galaxy's life-time (top panel). The stellar
metallicities are generally relatively low, with a median value of $\sim
0.3\,{\rm Z}_{\sun}$, with larger values associated to larger galaxies (see
below). The results illustrated in Fig.~\ref{fig:accretedGA3} are in good
agreement with those by \citet{Font_etal_2006} who find that one or a few more
satellites in the range $10^8-10^{10}\,{\rm M}_{\odot}$ can make up 50-80 per
cent of the stellar halo and most of them are accreted early on (${\rm t}_{\rm
  accr} > 9$~Gyr).

\begin{figure*}
\bc
\resizebox{7.5cm}{!}{\includegraphics[]{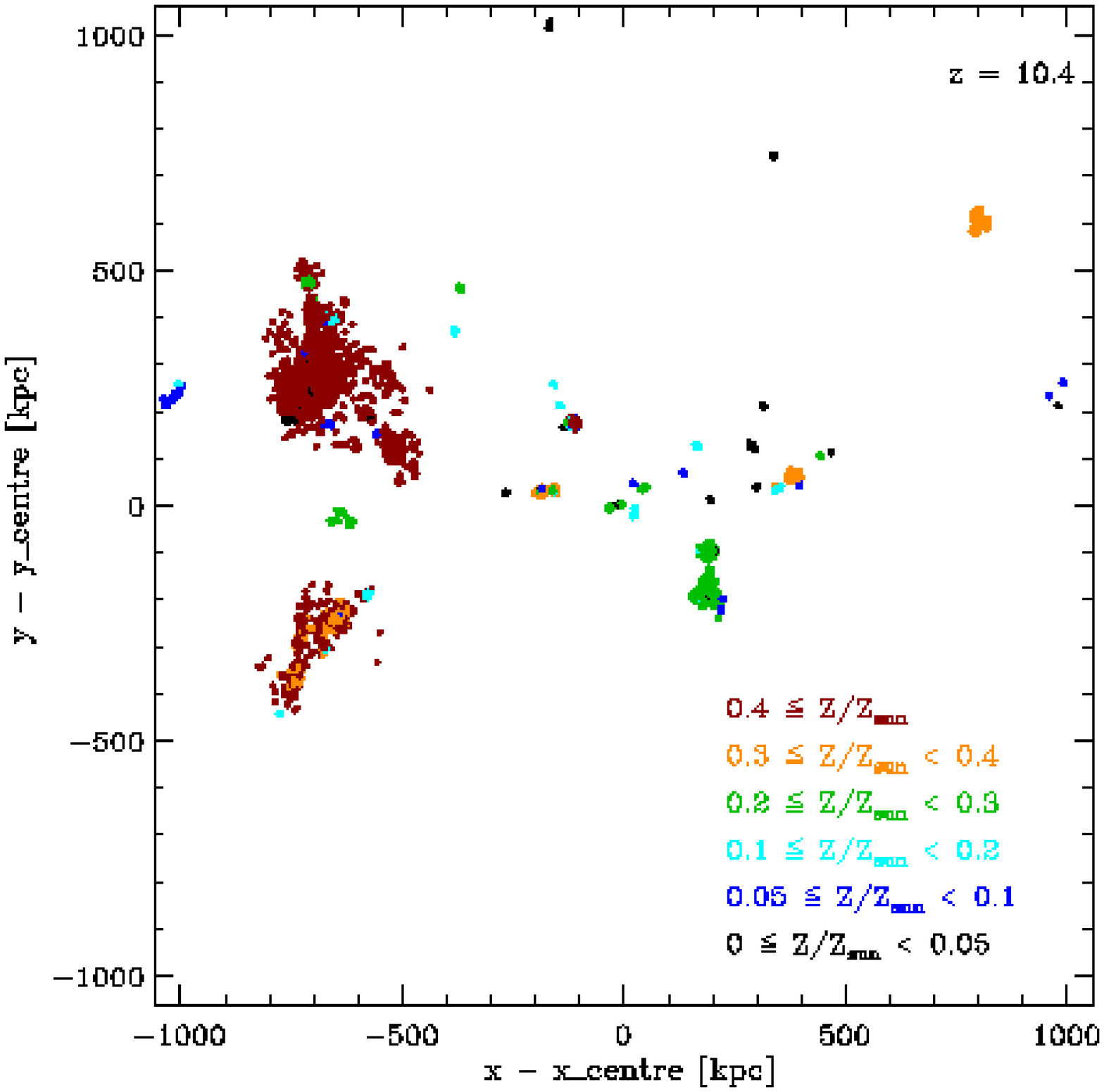}}
\resizebox{7.5cm}{!}{\includegraphics[]{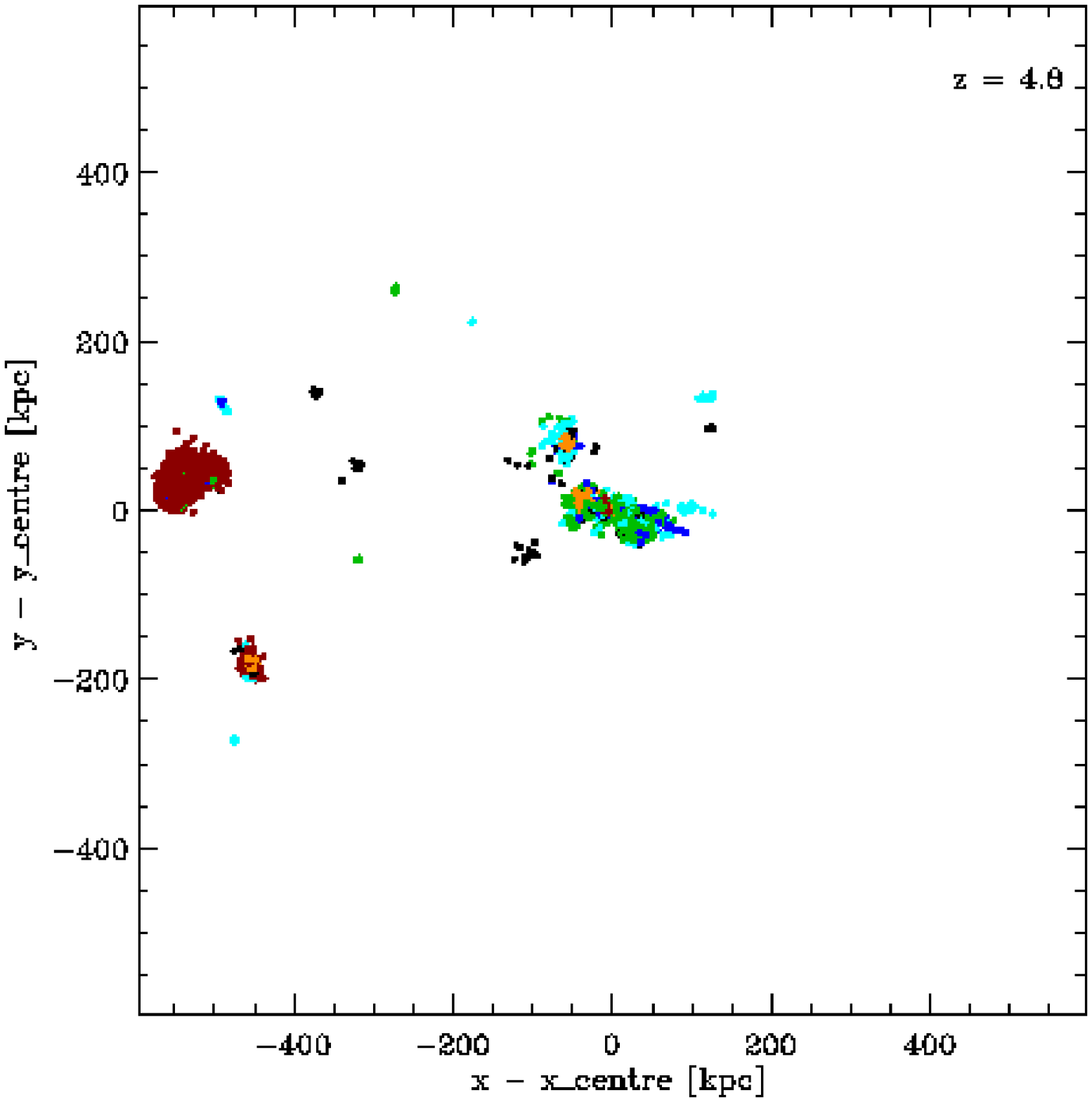}}
\resizebox{7.5cm}{!}{\includegraphics[]{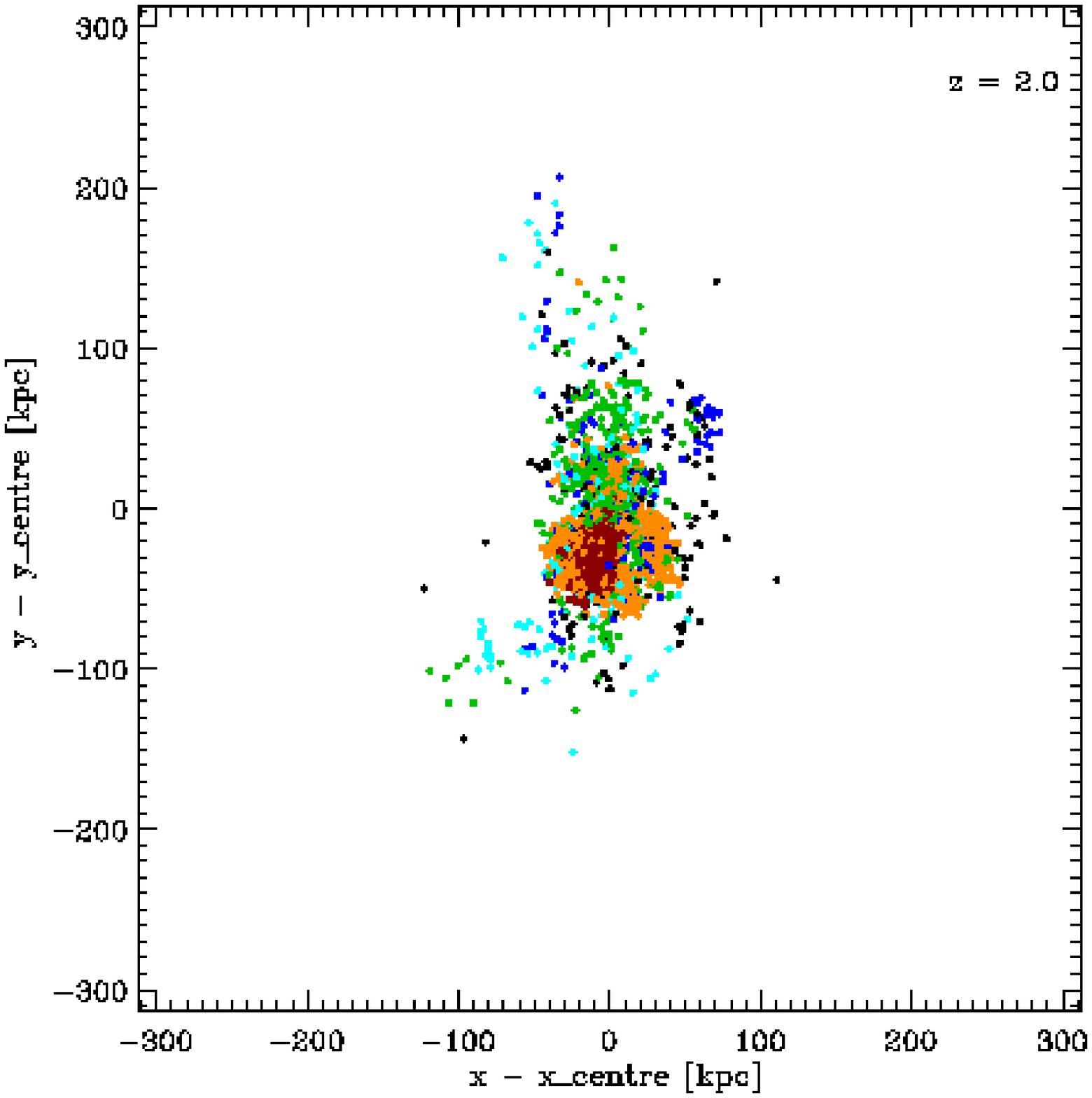}}
\resizebox{7.5cm}{!}{\includegraphics[]{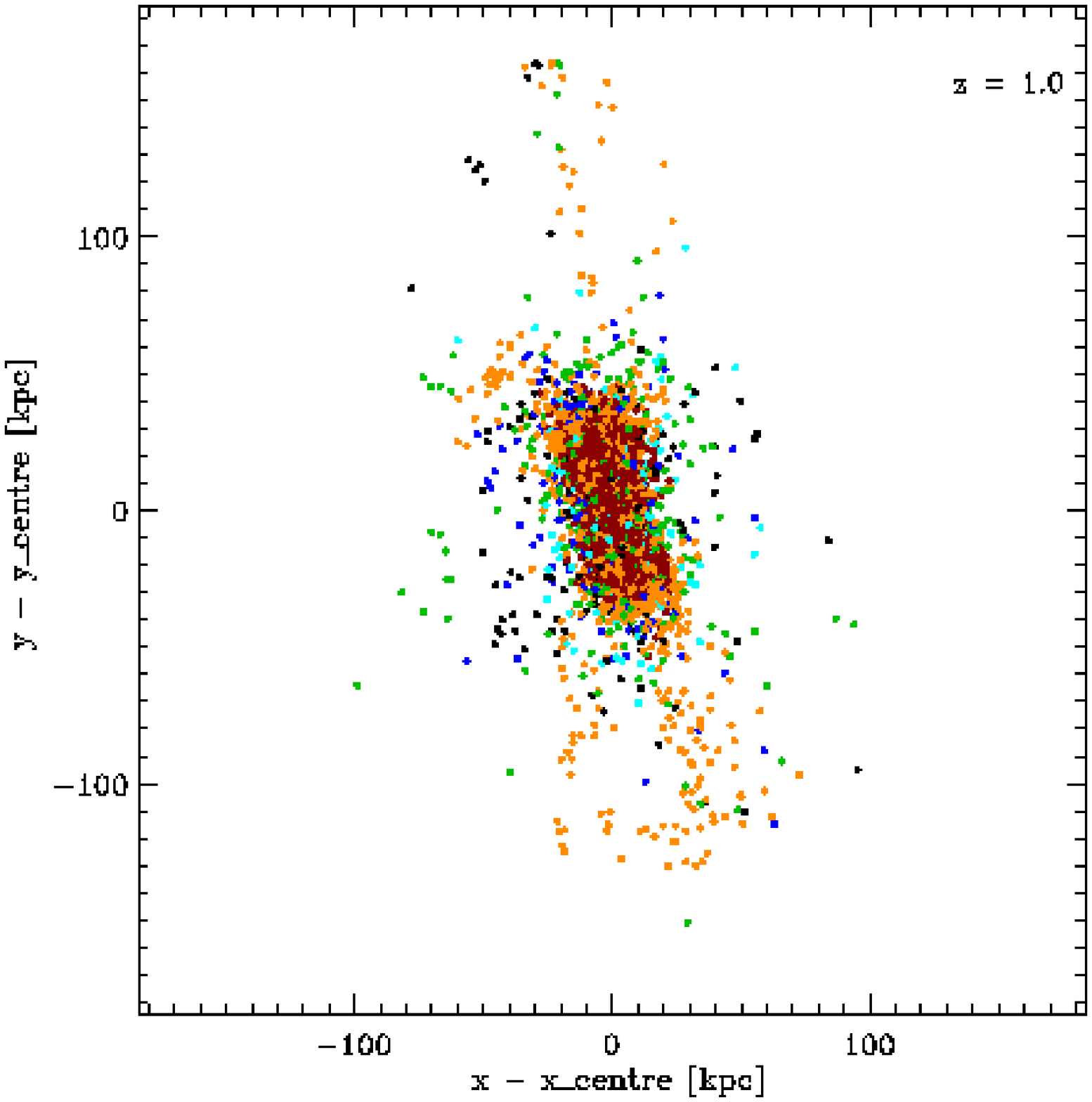}}
\resizebox{7.5cm}{!}{\includegraphics[]{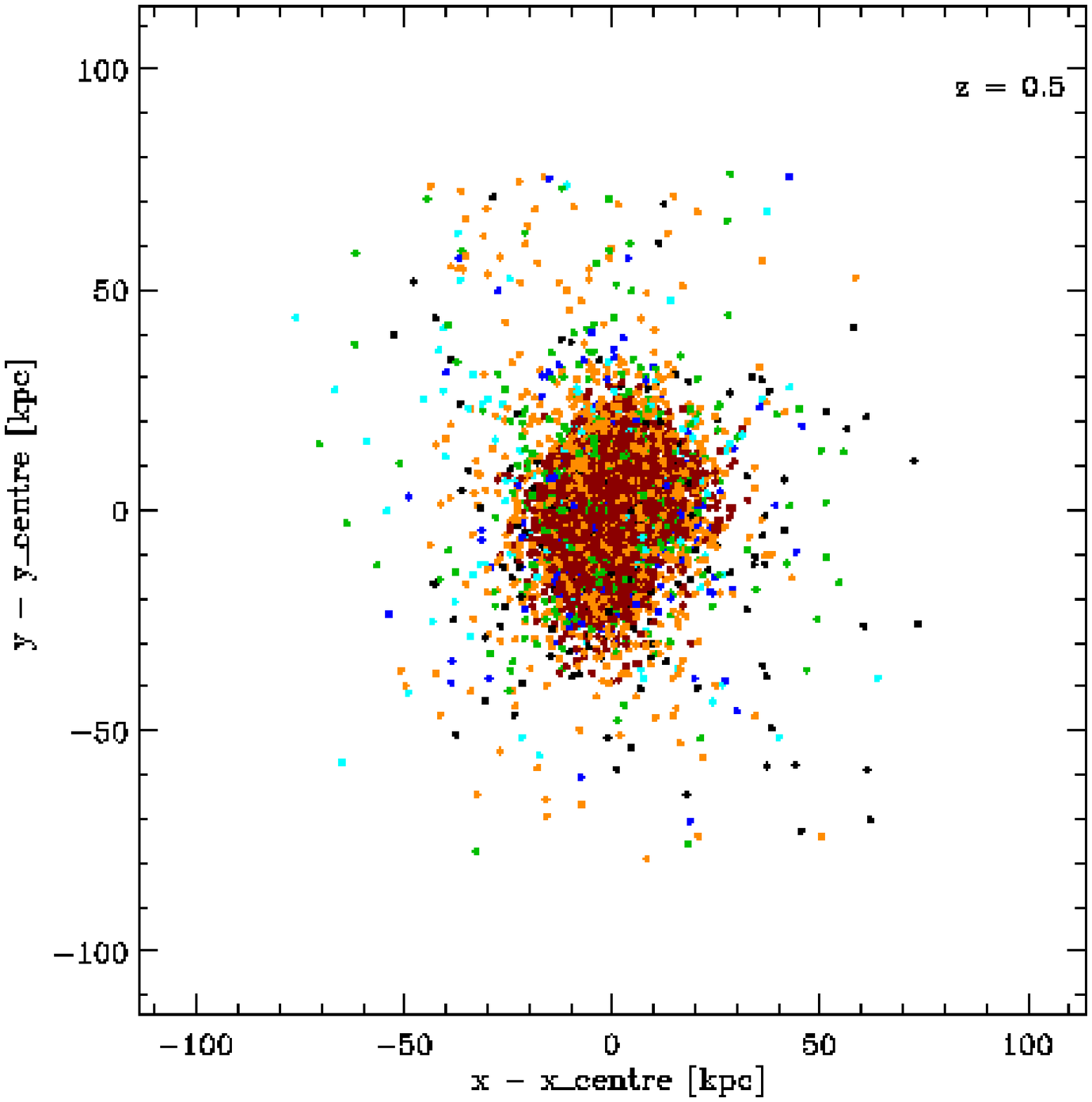}}
\resizebox{7.5cm}{!}{\includegraphics[]{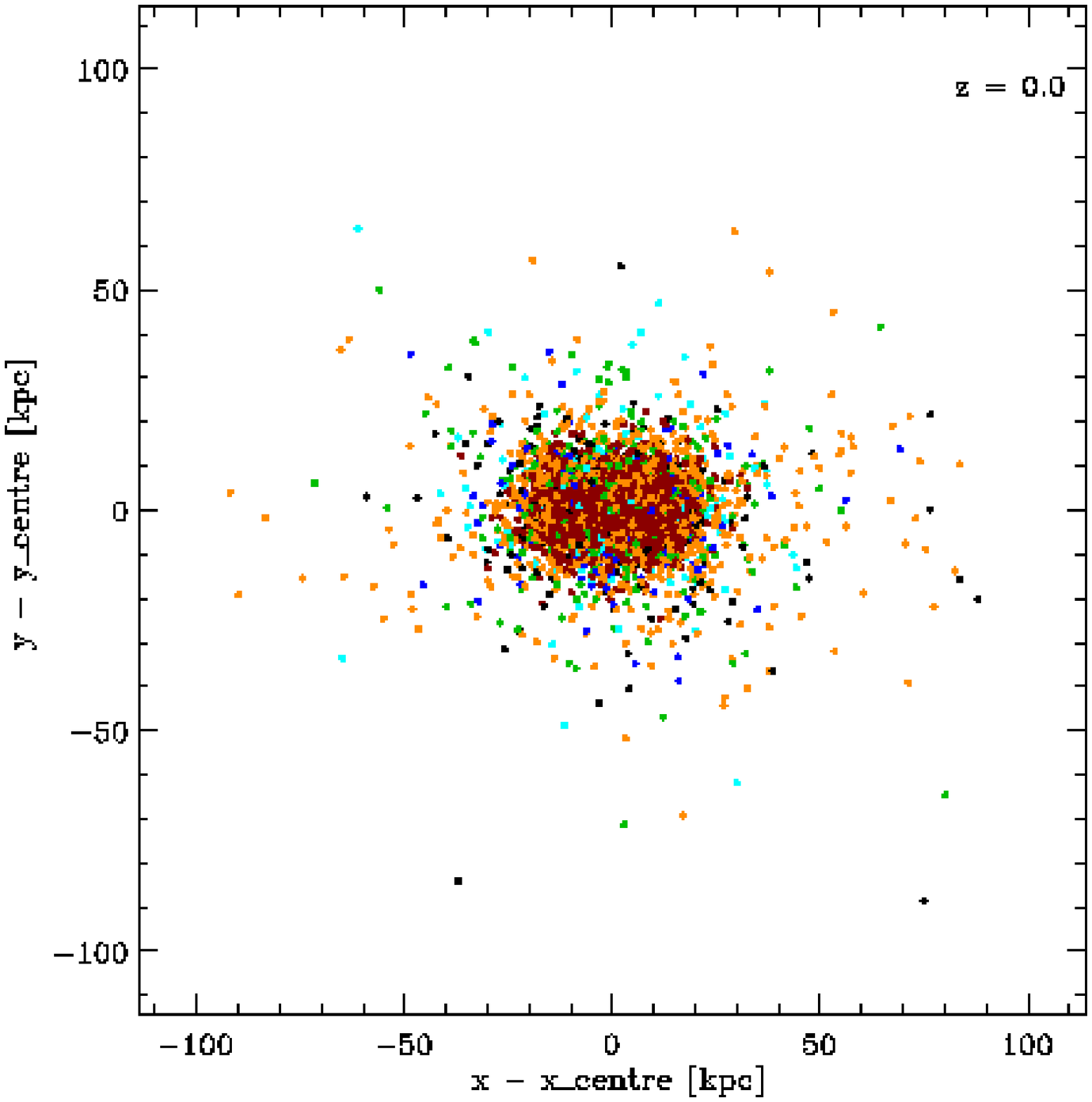}}
\caption{Projected distribution of the star particles that end up in
  the stellar halo of the highest resolution simulation used in our study
  (GA3), at the same redshifts as in Fig.~\ref{fig:GA3maps}, and colour-coded
  as a function of their metallicity as indicated in the top left panel. If we
  assume a [$\alpha$/Fe]$\sim +0.4$~dex, the highest metallicity bin could be
  translated into $-0.7\lesssim$~[Fe/H], while the lowest to
  [Fe/H]$\lesssim-1.6$ \citep{salaris}. As for Fig.~\ref{fig:GA3maps}, the box
  at $z=0$ is centred on the most bound particle of the GA3 halo, while those
  at higher redshifts are centred on the most bound particle of the main
  progenitor at the corresponding redshift. }
\label{fig:starpart}
\ec
\end{figure*}

Fig.~\ref{fig:starpart} shows the projected distribution of the star particles
that end up in the stellar halo of the highest resolution simulation used in
our study (GA3). For this figure, and for the results presented in the
following, we use 10 per cent of the most bound particles of every accreted
halo, but similar results are obtained if 5 per cent of the particles are
selected. We obtain 4188 star particles for the simulation GA2, and 31716 star
particles for the simulation GA3.  The star particles in
Fig.~\ref{fig:starpart} are colour coded as a function of their metallicity, as
indicated in the top-left panel, and their spatial distribution is shown at the
same redshifts used for Fig.~\ref{fig:GA3maps}.  The star particles that end up
in the stellar halo extend over a projected region of $\sim 1\,{\rm Mpc}^2$
comoving at redshift $\sim 10$. At redshift $\sim 1$, the star particles are
already assembled in a single relatively elongated component which becomes
progressively more spherical with decreasing redshift (see below). We note that
the the `stellar halo' of the simulation GA3 is dominated by particles
contributed by a single object that merged $\sim 7$~Gyr ago (and whose dark
halo was accreted 11 Gyr ago) and with metallicity $0.53\,{\rm Z}_{\sun}$ (see
Fig.~\ref{fig:accretedGA3}). This single galaxy contributes 19921 star
particles (i.e. about 60 per cent of all the star particles in the stellar
halo).

The lower panels of Fig.~\ref{fig:starpart} seems to suggest a slight
concentration of the most metal-rich stars, but no clear correlation between
metallicity and distance, with low metallicity and high metallicity stars
distributed all at various distances. It is interesting that in some cases
there is a clear difference between the mean metallicity of stars in a given
satellite (pockets of stars of different colours) - which is the basis of the
`chemical tagging' argument discussed in \citet{Freeman_Bland-Hawthorn_2002}.
The resolution of the simulations used in this study is, however, too low to
see any spatially coherent stellar streams in the halo at the present day.

\begin{figure}
\bc
\hspace{-0.8cm}
\resizebox{8cm}{!}{\includegraphics[]{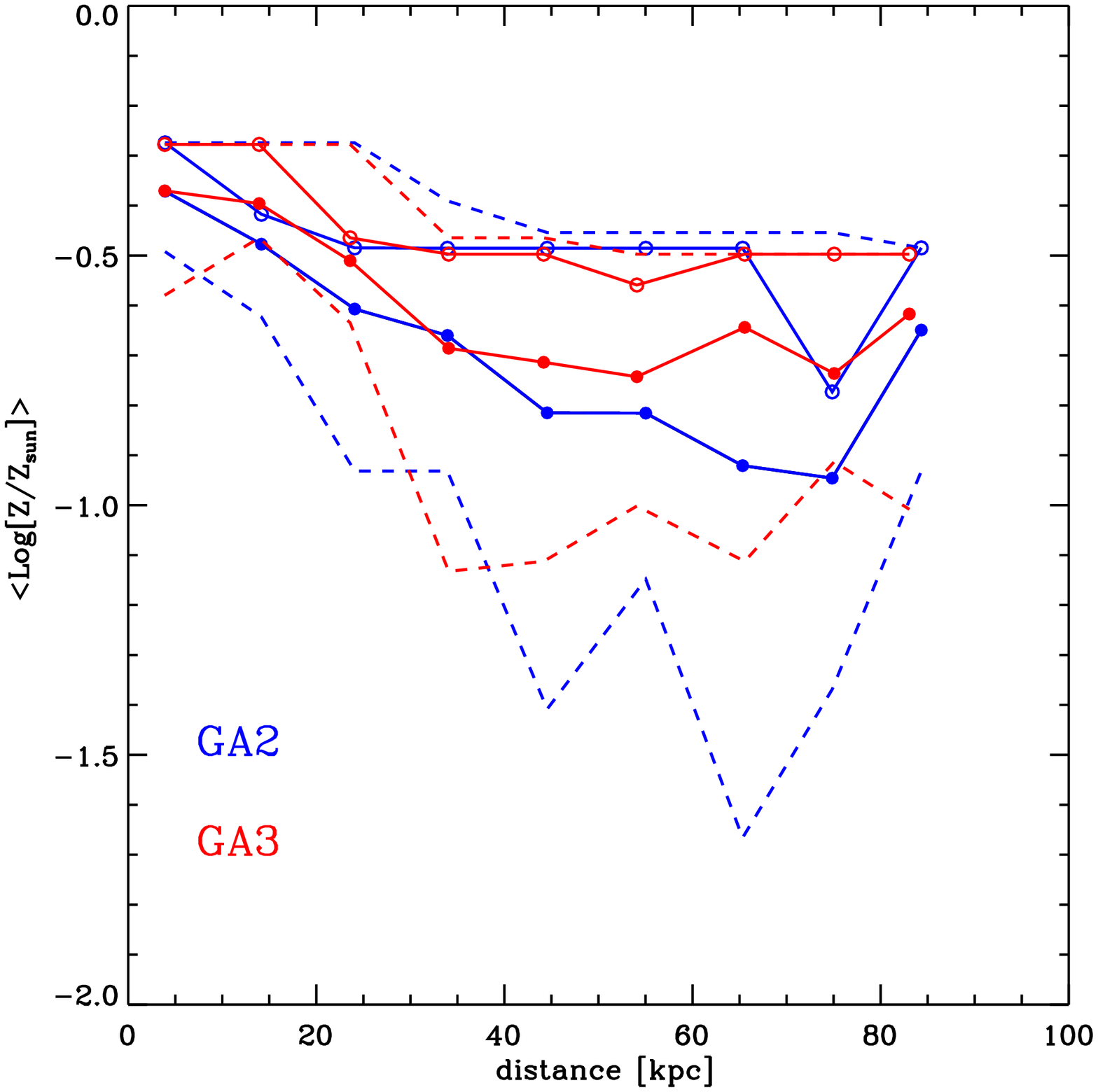}}
\caption{Mean (filled circles) and median (empty circles) metallicity of star
  particles as a function of the distance from the most bound particle in the
  Milky Way halo for the simulations GA2 (blue) and GA3 (red). Dashed
  lines correspond to the 15th and 85th percentiles of the distributions.}
\label{fig:halo_gradient}
\ec
\end{figure}

\begin{figure}
\bc
\hspace{-0.7cm}
\resizebox{8cm}{!}{\includegraphics[]{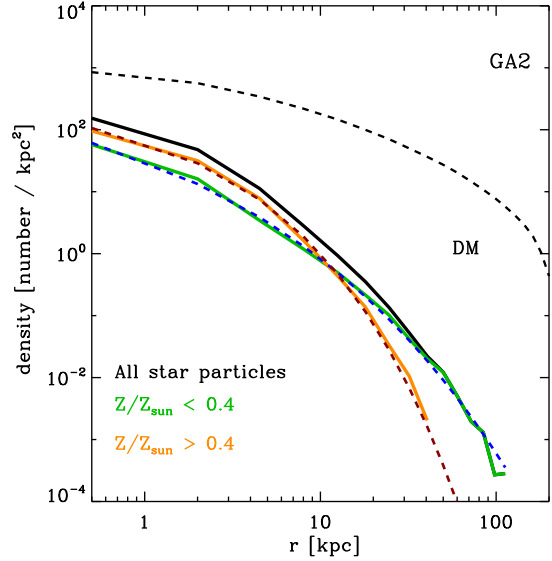}}

\hspace{-0.7cm}
\resizebox{8cm}{!}{\includegraphics[]{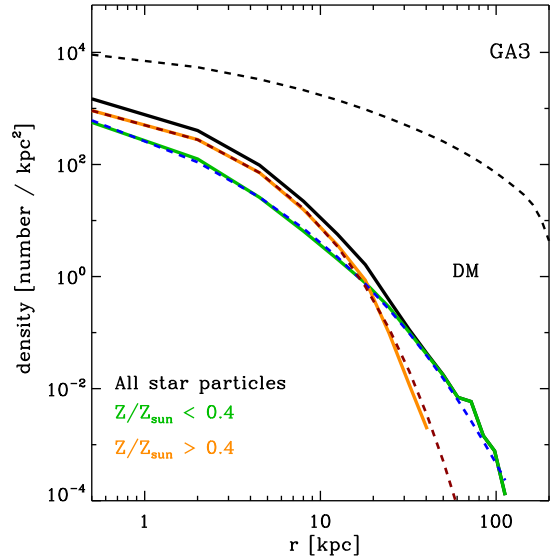}}
\caption{Projected density profile of the stellar halo (solid black lines) and
  of the dark matter halo (dashed black lines) for the simulations GA2 (top
  panel) and GA3 (bottom panel). The solid green and orange lines show the
  projected density profiles for star particles with metallicity smaller and
  larger than $0.4\,Z_\odot$ respectively.  The blue and red dashed lines show
  S\'ersic fits to the profiles of star particles in the two metallicity bins.}
\label{fig:profiles}
\ec
\end{figure}

The absence of a clear metallicity gradient\footnote{We have verified that an
  age gradient is also absent.} is illustrated more explicitly in
Fig.~\ref{fig:halo_gradient} which shows the metallicity of the star particles
as a function of the distance from the most bound particle in the Milky Way
halo for the simulations GA2 (blue) and GA3 (red). Filled and empty circles
connected by solid lines indicate the mean and the median of the distributions
respectively. Dashed lines correspond to the 15th and 85th percentiles. In both
simulations, the mean metallicity decreases from ${\rm
  Log[Z/Z}_{\sun}{\rm]}\sim -0.4$ at the centre to $\sim -0.8$ at a distance of
$\sim 40\,{\rm kpc}$. The median and upper 85th percentile of both
distributions are approximately flat around $\sim -0.5$. The lower 15th
percentile exhibits a certain decline with increasing distance from the centre,
suggesting that the inner region is largely dominated by high-metallicity stars
while the contribution from lower metallicity stars becomes more important
moving to the outer regions. We recall, however, that both distributions are
dominated in number by star particles associated to one or a few accreted
galaxies with relatively high metallicity (hence the flat behaviour of the
median and upper percentile): 58 per cent of the stellar halo particles in GA2
have metallicity $\geq -0.4$, and this fraction rises to 67 per cent for the
simulation GA3. As discussed previously, the metallicity of our stellar halo is
higher than what is known for the Galactic halo near the Sun. This could be
reflecting that the metallicity of the accreted galaxies may be too high (see
below).

From the observational viewpoint, it is interesting to ask where the most metal
poor stars of the stellar halo are located. We address this question in
Fig.~\ref{fig:profiles} which shows the projected density profiles (solid black
lines) of the star particles for the simulations GA2 (top panel) and GA3
(bottom panel) obtained by stacking the three projections on the xy, xz, and yz
plane. The dashed black lines in Fig.~\ref{fig:profiles} show the projected
density profile of the dark matter halo\footnote{We use only the particles in
  the `main halo', i.e. the self-bound part of the friends-of-friends halo.}.
In agreement with previous findings \citep{Bullock_Johnston_2005}, we find that
the profile of the stellar halo is steeper and more centrally concentrated than
the dark matter profile. The half-mass radius of the stellar halo in GA3 is
located at $\sim 4.3$ kpc, i.e. well within the Solar circle, as estimated for
the Milky Way \citep{Frenk_White_1982}. The solid orange and green lines show
the projected profiles of star particles with metallicity larger and smaller
than $0.4\,Z_\odot$ respectively. Fig.~\ref{fig:profiles} shows that star
particles with metallicity larger than $0.4\,Z_\odot$ are more centrally
concentrated than star particles of lower abundances. This implies that the
probability of observing low-metallicity stars increases if one looks at larger
distances from the Galactic centre ($\gtrsim 10-20$~kpc) where the contribution
from the inner more metal-rich star particles is less dominant.

Interestingly, our results appear to be in qualitative agreement with the
measurements by \citet{Carollo_etal_2007}. These authors have analysed a large
sample of over 20,000 stellar spectra from SDSS, and demonstrated that the
stellar halo of the Milky Way can be considered as the superposition of two
components which are spatially, kinematically, and chemically distinct. One
component, which they call `inner halo', dominates the population of stars
found at distances up to 10-15 kpc from the Galactic centre and peaks at 
higher metallicity than the `outer halo', which dominates in regions beyond
15-20 kpc.

\begin{figure}
\bc
\hspace{-0.7cm}
\resizebox{8cm}{!}{\includegraphics[]{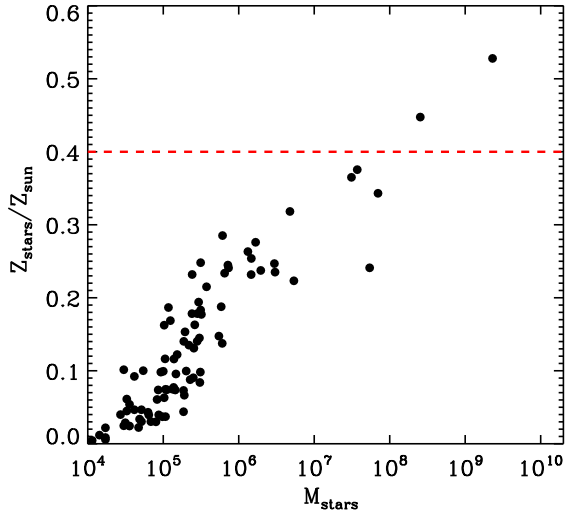}}
\caption{Stellar metallicity as a function of stellar mass for the 
  galaxies contributing to the stellar halo in the simulation GA3.}
\label{fig:met_mass}
\ec
\end{figure}

In order to understand this `duality' we turn to our simulations. From the
middle panel of Fig.~\ref{fig:accretedGA3} we see that stellar metallicity does
not correlate with the time of the merger of the galaxies that contribute to
the stellar halo. Rather, the `duality' of our model stellar halo originates
from a correlation between the stellar metallicity and the mass in stars of the
accreted galaxies, as shown in Fig.~\ref{fig:met_mass}. Since the most massive
galaxies can decay through dynamical friction to the inner regions of the host
halo, this is where higher metallicity stars will be found preferentially. Note
that the mass-metallicity relation in our simulations appears to be offset from
that observed in the sense that the accreted galaxies are too metal-rich,
explaining why our stellar halo also has too high a metallicity. This might be
due to a too efficient mixing of metals with the intergalactic medium (we
recall that we are using a 100 per cent mixing efficiency), and/or to
inefficient feedback in the building blocks of the stellar halo. We will
analyse this in further detail in a future paper where we will study the
satellite population of our model Milky Way.

A large fraction of galaxies accreted onto the main branch contribute to the
star particles in the inner halo\footnote{70 and 95 per cent of the accreted
  galaxies contribute at least one particle to this region of the stellar halo
  for the simulations GA2 and GA3 respectively.}, i.e $r < 10$ kpc, but only a
few systems contribute significantly to the population of star particles in
this region.  For both simulations, about 65 per cent of the star particles in
the inner region comes from one single system that merged onto the main branch
$\sim 8$~Gyr ago for the simulation GA2, and $\sim 7$~Gyr ago for the
simulation GA3. Another $\sim 20$ per cent of the star particles in the inner
10 kpc are contributed in approximately equal fractions from other three
galaxies that merged between $\sim 8$ and $\sim 10.5$~Gyr ago for the
simulation GA2, and a bit earlier (between $\sim 10$ and $\sim 11.5$~Gyr ago)
for the simulation GA3.

The blue and red dashed lines in Fig.~\ref{fig:profiles} show S\'ersic fits to
the star particles with metallicity larger and smaller than $0.4\,Z_\odot$
respectively (corresponding to the orange and green lines). The fitted function
is given by:
\begin{displaymath}
  \Sigma (R) = \Sigma_{\rm eff} \exp\left[ -b_n[(R/R_{\rm
  eff})^{(1/n)} - 1]\right]
\end{displaymath} 
where $R_{\rm eff}$ is the radius containing half the light and $\Sigma_{\rm
  eff}$ is the surface brightness at that radius. The results of the fit are
given in Table~\ref{tab:fitresults}. For both simulations, the profile of star
particles with metallicity lower than $0.4\,{\rm Z}_{\sun}$ is well fit by a
S\'ersic profile with index $n \sim 3$ and half-light radius $\sim 4$~kpc. For
the star particles with larger metallicity the characteristic radius is
comparable, and the S\'ersic index is smaller (the distribution falls off more
steeply). The logarithmic slope of the corresponding density profile at $R_{\rm
  eff}$ is $\gamma \sim -3.3$ for the metal-rich component and $\gamma \sim
-3.1$ for the metal-poor stars, in very good agreement with those measured for
the stellar halo of the Milky Way at a similar galactocentric distance.

\begin{table}
\caption{Results of the S\'ersic fit (see text) to the projected profiles of
  star particles with metallicity larger and smaller than $0.4\,{\rm Z}_{\sun}$
  (orange and green lines in Fig.~\ref{fig:profiles}).}   
\begin{tabular}{lll}
  \hline
  ${\rm Z} < 0.4\,{\rm Z}_{\sun}$ & $R_{\rm eff}$ [kpc] & $n$ \\
  \hline
  GA2 & $4.1$ & $3.3$\\
  
  GA3 & $4.9$ & $3.0$\\ \hline ${\rm Z} \geq 0.4\,{\rm Z}_{\sun}$ &
  $R_{\rm eff}$ [kpc] & $n$ \\ \hline GA2 & $4.7$ & $1.9$\\
  
  GA3 & $4.4$ & $1.6$\\
  \hline

\end{tabular}
\label{tab:fitresults}
\end{table}

\begin{figure}
  \bc \hspace{-0.6cm} \resizebox{9cm}{!}{\includegraphics[bb= 16 120 593
    660,clip]{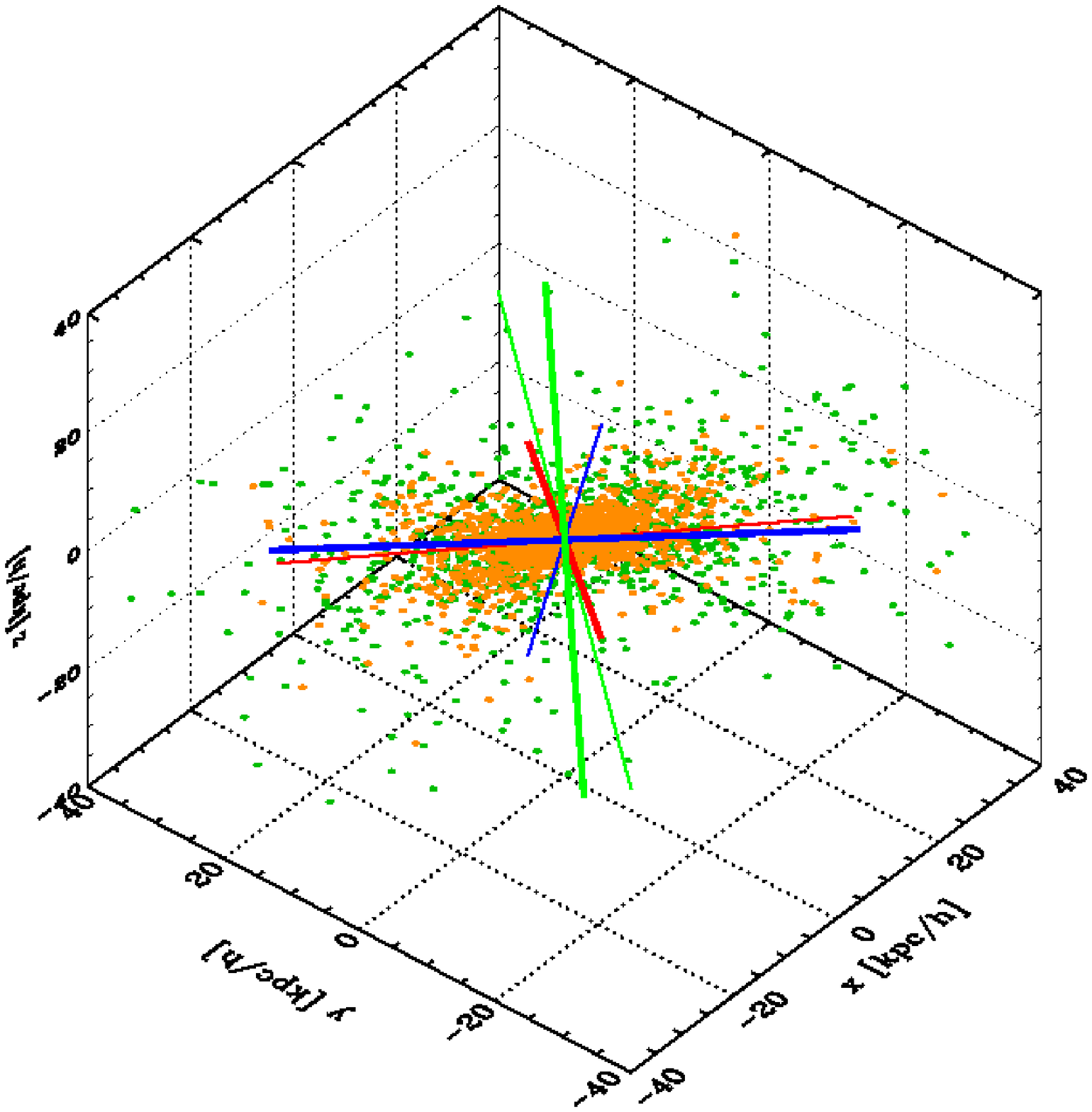}}

\hspace{-0.6cm}
\resizebox{9cm}{!}{\includegraphics[bb= 16 120 593 660,clip]{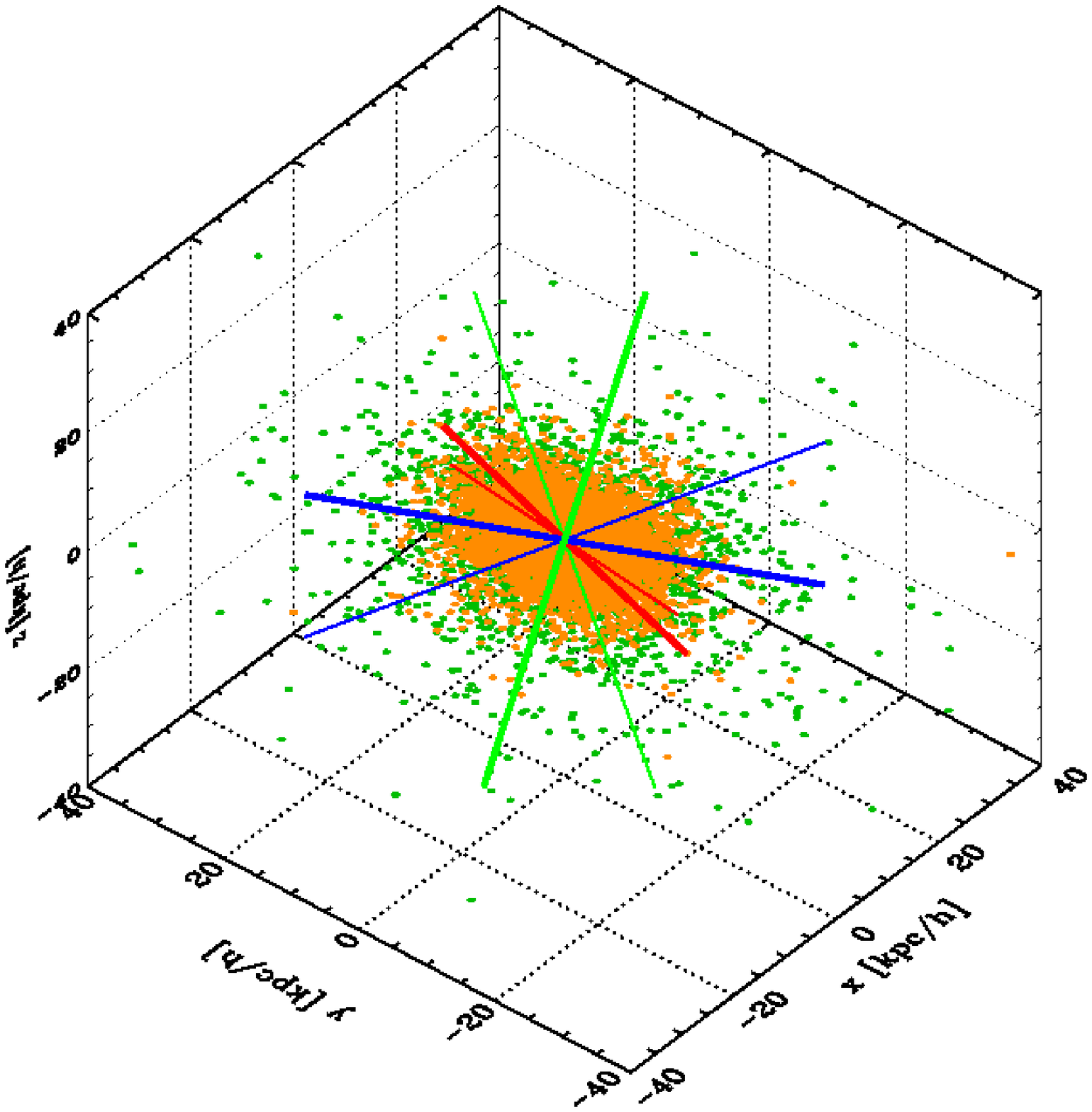}}
\caption{Three-dimensional distribution of the star particles from the
  simulations GA2 (top panel) and GA3 (bottom panel). Green symbols refer to
  particles with ${\rm Z}/{\rm Z}_{\odot} < 0.4 $, while orange symbols refer
  to star particles with ${\rm Z}/{\rm Z}_{\odot} \geq 0.4 $. Red, blue and
  green lines represent the principal axes of the stellar halo (thin lines) and
  of the dark matter halo (thick lines). Red is used for the long axis, blue
  for the intermediate, and green for the short axis.}
\label{fig:axes}
\ec
\end{figure}

In order to characterise the three-dimensional shape of the stellar halo, we
have assumed that its isodensity surfaces can be approximated by triaxial
ellipsoids of the form:
\begin{displaymath}
  \frac{X^2_1}{a^2}+\frac{X^2_2}{b^2}+\frac{X^2_3}{c^2} = 1\,\,\, {\rm with}\,
  a \geq b \geq c
\end{displaymath}
where $a, b, {\rm and}\, c$ are the lengths of the three axes and $X_{\alpha}$
(with $\alpha = 1, 2, 3$) is the coordinate with respect to the $\alpha$ axis
and relative to the position of the most bound particle of the Milky Way halo.
The directions and lengths of the major axes can then be computed by finding
the eigenvectors and eigenvalues of the matrix $M_{\alpha\beta}$:
\begin{displaymath}
  M_{\alpha\beta} = \sum_i X^i_{\alpha} X^i_{\beta}
\end{displaymath}
where $\alpha{\rm ,}\,\beta = 1, 2, 3$. In Table~\ref{tab:axes}, we list the
resulting axial ratios $c/a$ and $c/b$ for the stellar halo and for the dark
matter halo, for the simulations GA2 and GA3.  The principal axes of the
stellar halo and of the dark matter halo are plotted as thin and thick solid
lines in Fig.~\ref{fig:axes}. The top panel corresponds to the simulation GA2,
while the bottom panel is for the simulations GA3. In both panels, orange and
green points are star particles with metallicity larger and smaller than
$0.4\,Z_{\sun}$

\begin{table}
\caption{Axial ratios for the stellar halo and for the dark matter halo of the
  Milky Way galaxy for the simulations GA2 and GA3 used in this study.}  
\begin{tabular}{lll}
  \hline
  stellar halo & $c/a$  & $c/b$ \\
  \hline
  GA2 & $0.51$ & $0.74$\\
  
  GA3 & $0.45$ & $0.89$\\
  \hline
  dark matter halo & $c/a$ & $c/b$ \\
  \hline
  GA2 & $0.70$ & $0.80$\\
  
  GA3 & $0.72$ & $0.85$\\
  \hline

\end{tabular}
\label{tab:axes}
\end{table}
The value of the short-to-long ($c/a$) axis ratio measured for the dark matter
halo is in good agreement with previous numerical studies in a $\Lambda$CDM
cosmology by \citet{Bullock_2002} who finds typical axis ratios to be $c/a \sim
0.6-0.8$ for Milky Way size haloes. The stellar halo has a short-to-long axis
ratio $\sim 0.45-0.5$, slightly lower than observational measurements
\citep[][and references therein]{Chiba_and_Beers_2000,Bell_etal_2007}.
Interestingly, for both simulations the minor axis of the stellar halo is quite
well aligned with the the minor axis of the dark matter halo: the angle between
these two axes is only $\sim 10$~deg for the simulation GA2 and $\sim 30$~deg
for the simulation GA3. For the highest resolution simulation (GA3), also the
major and intermediate axes of the stellar halo are well aligned with the
corresponding axes of the dark matter halo (the angular distance is $\sim 9$
and $\sim 30$~deg for the other two axes). For the simulation GA2, the angular
distance between the major and intermediate axes of the stellar halo and the
corresponding axes of the main halo is $\sim 75$~deg.

\section{Discussion and conclusions}
\label{sec:discconcl}

During the last decade, a number of observational tests of the standard
cosmological model have ushered in a new era of `precision cosmology'. While
the basic theoretical paradigm for structure formation is now well established,
our understanding of the physical processes governing galaxy formation and
evolution is far from complete. Modern advances in ground- and space-based
observational capabilities have enabled us to study the resolved stellar
populations of Local Group galaxies (and beyond) in unprecedented detail. For
our own Galaxy and for a number of the brightest members of the Local Group, a
wealth of observational data is now available about the ages and chemical
abundances of their stars. Much more data will become available over the next
decade, providing important tests for current models of galaxy formation and
evolution.

In this paper, we have discussed the formation of the Milky Way and of its
stellar halo in the context of a hybrid cosmological approach that combines
high-resolution simulations of a `Milky-Way' halo with semi-analytic methods.
  
Our approach is similar to that adopted by \citet{Bullock_Johnston_2005}, but
the two methods differ in a number of details.
\citeauthor{Bullock_Johnston_2005} have used mass accretion histories of Milky
Way-size galaxies using the extended Press-Schechter formalism. For each
accretion event in the analytic merger trees, they have run $N$-body
simulations following the dynamical evolution of the accreted dark matter
satellite in an analytic parent galaxy+host halo potential. The stellar
distribution and properties of each satellite are constructed using a variable
mass-to-light ratio to each dark matter particle \citep{Font_etal_2006}, and
the chemical evolution of the satellites is modelled taking into account the
enrichment from both Type II and Type Ia supernovae
\citep{Robertson_etal_2005}. In our work, we have used a set of increasing
resolution $N$-body simulations of a Milky Way-like halo, combined with a
semi-analytic method that allows us to model the stellar distribution of the
galaxies self-consistently during the N-body simulation. While our approach is
based on a fully numerical simulation, the numerical resolution for each
accreted satellite is lower than that achieved by
\citeauthor{Bullock_Johnston_2005}. In addition, our chemical evolution model
assumes an instantaneous recycling approximation which is appropriate for
elements produced by Type II supernovae, but wrong for the iron-peak elements
which are mainly produced by Type Ia supernovae.
  
The galaxy formation model employed in this work has been studied in a number
of previous papers, and it has been shown to successfully reproduce a number of
observational results for the global galaxy population in the local Universe
and at higher redshifts. The physical properties of our model Milky-Way
galaxies are in quite nice agreement with the observational results, and the
predicted evolutions are very similar, over the entire numerical range covered
by the simulations used in our work (from $\sim 2\times10^8$ to $\sim
3\times10^5\,{\rm M}_{\sun}$ dark matter particle mass).  This is the first
time to our knowledge that model convergence has been shown on a
galaxy-by-galaxy basis\footnote{\citet{Springel_etal_2001} discussed the
convergence of model results in statistical sense using a set of four
high-resolution $N$-body simulations of galaxy clusters.}.
  
Our model Milky Way galaxy is a relatively young system (50 per cent of the the
stars are assembled in a single object only at redshift $\sim 0.8$) with a very
old spheroidal component (50 per cent of the stars in the spheroid are in place
at redshift $\sim 3$), which is formed through a series of minor mergers and a
few episodes of disk instability occurring early on during the galaxy's
lifetime. All stars in the spheroidal component are old ($\gtrsim 11$~Gyr)
while the stars in the disk have a much larger spread in age, reflecting a
prolonged star formation activity which is in qualitative agreement with
observational determinations. A detailed comparison between model and observed
metallicity distributions is complicated by the use of the instantaneous
recycling approximation, which does not allow direct comparison with the
observed iron distribution. Using the observed relation between oxygen and iron
abundances, we obtain a metallicity distribution which is similar to the
observed one for the disk component, but a spheroidal component which is less
enriched than the observed Galactic bulge (although small changes in the
parameters of the semi-analytic models are able to remove this discrepancy
without significantly affecting the remaining properties of the simulated
Galaxy, see e.g Fig.~\ref{fig:grid}).
  
Assuming that the stellar halo builds up from the cores of the satellite
galaxies that merged with the Milky Way over its life-time, we are able to
study the physical and structural properties of this Galactic component. Our
model stellar halo is made up of very old stars (the majority of the stars
formed earlier than 10 Gyr ago) with low metallicity ($\lesssim 0.5\,{\rm
  Z}_{\sun}$), although relatively high in comparison to the Galactic halo near
the Sun. A few relatively massive ($10^8-10^{10}\,{\rm M}_{\sun}$) satellites
accreted early ($> 9$~Gyr) contribute the largest fraction of the star
particles that end up in the stellar halo, in agreement with previous results
by \citet{Font_etal_2006}. There is no evidence of a metallicity gradient for
halo stars, but we find evidence for a stronger concentration of higher
metallicity stars.  This implies that the probability of observing
low-metallicity halo stars increases with distance from the Galactic centre
($\gtrsim 20$~kpc). The `duality' we find for our model stellar halo does not
originate from a correlation between metallicity and accretion time, rather
from the fact that the main contributors to the stellar halo (satellites with
larger mass) have preferentially higher metallicity than lower mass systems. We
find that the three-dimensional distribution of halo stars is well described by
a triaxial ellipsoid with short-to-long axis $\sim 0.5-0.6$, in agreement with
observational measurements, and whose axes are well aligned to those of the
parent dark matter halo.

A more accurate treatment of chemical enrichment will allow us to carry out a
more detailed comparison with observed chemical compositions, and to establish
similarities and differences between present-day satellites and the Galactic
building blocks. The numerical resolution of the simulations used in this work,
however, is too low for studies of spatially and kinematically coherent stellar
streams in the present day stellar halo. Higher resolution simulations are
therefore needed for these studies. These are all much needed steps to
interpret the outcome of large surveys such as SEGUE, RAVE and ultimately Gaia,
with the goal of unrevealing the evolutionary history of our Galaxy.


\section*{Acknowledgements}
We are indebted to Felix Stoehr for making his GA series available, and to
Volker Springel for making available the substructure finder and merger tree
construction software that was originally developed for the Millennium
Simulation project (http://www.mpa-garching.mpg.de/galform/virgo/millennium/).
We thank K. Dolag for help in adapting the software to the simulation outputs,
M. Zoccali for providing observational measurements prior to publication, and
Y.S. Li for useful discussions. We thank Simon White for encouragement in
completing this project.  GDL acknowledges the hospitality of the Kapteyn
Astronomical Institute of Groningen, where this project was initiated.  AH
gratefully acknowledges financial support from the Netherlands Organisation for
Scientific Research (NWO) and from the Netherlands Research School for
Astronomy (NOVA).  GDL thanks D. Gadotti for useful discussions about bar
formation and evolution, U.  Maio for refreshing discussions about eigenvalues
and eigenvectors, and A.  Moretti for useful comments on a preliminary version
of this paper, as well as for continuous support in the last year.

\bsp

\label{lastpage}

\bibliographystyle{mn2e}
\bibliography{mw_delucia_rev}

\end{document}